\documentclass[acmsmall]{acmart}

\setcopyright{cc}
\setcctype{by}
\acmDOI{10.1145/3808129}
\acmYear{2026}
\acmJournal{PACMSE}
\acmVolume{3}
\acmNumber{FSE}
\acmArticle{FSE122}
\acmMonth{7}
\acmSubmissionID{fse26mainb-p337-p}
\received{2026-02-24}
\received[accepted]{2026-03-24}

\usepackage{xspace}
\usepackage{enumitem}
\usepackage{multirow}
\usepackage{rotating}
\usepackage{pifont}
\usepackage{tcolorbox}
\usepackage{subcaption}
\usepackage{tikz}
\usepackage{ifthen}
\usepackage{xcolor}
\usepackage{listings}
\usepackage{multicol}
\usepackage{array}
\usepackage{makecell}

\usetikzlibrary{calc}

\definecolor{diffadded}{RGB}{0,128,0}
\definecolor{diffdeleted}{RGB}{255,0,0}

\lstdefinelanguage{diff}{
    basicstyle=\ttfamily\footnotesize,
    morecomment=[f][\color{diffdeleted}]{-}, %
    morecomment=[f][\color{diffadded}]{+},   %
    morecomment=[f][\color{gray}]{@@},      %
}

\definecolor{mycolor}{rgb}{0.122, 0.435, 0.698}%
\definecolor{gray1}{gray}{0.3}

\definecolor{darkgreen}{rgb}{0.0, 0.5, 0.0}
\definecolor{darkred}{rgb}{0.82, 0.1, 0.26}

\newcommand{\result}[1]{%
\begin{tcolorbox}[colframe=mycolor,boxrule=0.5pt,arc=4pt,
      left=6pt,right=6pt,top=6pt,bottom=6pt,boxsep=0pt,width=\textwidth]%
      {#1}
\end{tcolorbox}%
}

\newcommand{\mysliderlen}{2em}

\def\SLIDER#1#2#3{%
\tikz[baseline=-0.1cm]{
 \coordinate (start) at (0,0);
 \coordinate (end) at (#1,0);
 \coordinate (mark) at ($(start)!#2!(end)$);
 \draw[line width=0.15cm, line cap=round, lightgray] (start) -- (end);

 \ifthenelse{\equal{#3}{circle}}{
  \ifdim #2pt=0pt
      \fill[draw=black!50!red, rounded corners=0.2mm, fill=red!80!white] (mark) circle(.12);
  \else
      \ifdim #2pt<0.34pt
        \fill[draw=black!50!orange, rounded corners=0.2mm, fill=orange!80!white] (mark) circle(.12);
      \else
        \ifdim #2pt<0.68pt
          \fill[draw=black!50!olive, rounded corners=0.2mm, fill=olive!80!white] (mark) circle(.12);
        \else
          \fill[draw=black!50!teal, rounded corners=0.2mm, fill=teal!80!white] (mark) circle(.12);
        \fi
      \fi
  \fi
  }
  {
    \ifdim #2pt=0pt
      \draw[draw=black!50!red, fill=red!80!white] (mark) ++(-0.13,0.0) -- ++(0.13,0.13) -- ++(0.13,-0.13) -- ++(-0.13,-0.13) -- cycle;
    \else
      \ifdim #2pt<0.34pt
        \draw[draw=black!50!orange, fill=orange!80!white] (mark) ++(-0.13,0.0) -- ++(0.13,0.13) -- ++(0.13,-0.13) -- ++(-0.13,-0.13) -- cycle;
      \else
        \ifdim #2pt<0.68pt
          \draw[draw=black!50!olive, fill=olive!80!white] (mark) ++(-0.13,0.0) -- ++(0.13,0.13) -- ++(0.13,-0.13) -- ++(-0.13,-0.13) -- cycle;
        \else
          \draw[draw=black!50!teal, fill=teal!80!white] (mark) ++(-0.13,0.0) -- ++(0.13,0.13) -- ++(0.13,-0.13) -- ++(-0.13,-0.13) -- cycle;
        \fi
      \fi
    \fi
  }
}
}

\newcommand{\toolname}{\textsc{Cleverest}\xspace}
\newcommand{\toolnamefuzz}{\textsc{ClevFuzz}\xspace}

\newcommand{\icse}[1]{{#1}}
\newcommand{\review}[1]{{#1}}

\newcommand{\gpt}{{\small{\texttt{GPT-4o}}}\xspace}
\newcommand{\gptmini}{{\small{\texttt{GPT-4o mini}}}\xspace}
\newcommand{\deepseekr}{{\small{\texttt{DeepSeek-R1}}}\xspace}

\newcommand{\xmark}{\makebox[.8em][c]{{\color{red}\ding{56}}\xspace}}
\newcommand{\xmarknospace}{\makebox[.8em][c]{{\color{red}\ding{56}}}}
\newcommand{\cmark}{\makebox[.8em][c]{{\color{teal}\ding{51}}\xspace}}
\newcommand{\cmarknospace}{\makebox[.8em][c]{{\color{teal}\ding{51}}}}
\newcommand{\rmark}{\makebox[.8em][c]{{\color{orange}\ding{227}}\xspace}}
\newcommand{\rmarknospace}{\makebox[.8em][c]{{\color{orange}\ding{227}}}}
\newcommand{\dmark}{\makebox[.8em][c]{{\color{olive}\ding{48}}\xspace}}
\newcommand{\dmarknospace}{\makebox[.8em][c]{{\color{olive}\ding{48}}}}

\newcommand{\mujs}{\textsf{Mujs}\xspace}
\newcommand{\libxml}{\textsf{Libxml2}\xspace}
\newcommand{\poppler}{\textsf{Poppler}\xspace}
\newcommand{\jerryscript}{\textsf{JerryScript}\xspace}
\newcommand{\php}{\textsf{PHP}\xspace}
\newcommand{\zthree}{\textsf{Z3}\xspace}
\newcommand{\jq}{\textsf{JQ}\xspace}
\newcommand{\micropython}{\textsf{MicroPython}\xspace}

\title{Evaluating LLM-Based Regression Test Generation}

\author{Jing Liu}
\authornote{Both authors contributed equally to this research.}
\orcid{0009-0008-1217-7467}
\affiliation{%
  \institution{MPI-SP}
  \country{Germany}
}
\email{jing6@acm.org}

\author{Seongmin Lee}
\authornotemark[1]
\orcid{0000-0003-0805-8947}
\affiliation{%
  \institution{University of California at Los Angeles}
  \country{USA}}
\email{seongminlee@sigsoft.org}

\author{Eleonora Losiouk}
\orcid{0000-0002-2315-7823}
\affiliation{%
  \institution{University of Padua}
  \country{Italy}}
\email{eleonora.losiouk@unipd.it}

\author{Marcel B{\"o}hme}
\orcid{0000-0002-4470-1824}
\affiliation{%
 \institution{MPI-SP}
 \country{Germany}}
\email{marcel.boehme@acm.org}

\begin{document}

\begin{abstract}
  Large Language Models (LLMs) have shown tremendous promise in automated software engineering.
  In this paper, we investigate the opportunities of LLMs for just-in-time regression test generation for programs, like parsers, interpreters, or compilers, that take highly structured, human-readable inputs. When a new bug fix or code change is committed, the repository (as part of the CI/CD workflow) runs an LLM for a few minutes to generate regression test cases for that commit that exercise the changed code and potentially trigger any bugs.

  Specifically, we investigate LLM-based regression test generation as a \emph{machine translation task} that takes the developer-provided commit message, the code change, and the name of the input format (e.g., XML) and produces regression test cases for the described change in the given input format.
  In our experiments testing \review{72} commits to \mujs, \libxml, \poppler, \jerryscript, \zthree, \php, \review{\jq, and \micropython}, our feedback-directed, zero-shot LLM-based prototype \toolname performed well, even if we did \emph{not} provide the code change. In under 2 minutes, on average, \toolname found as many bugs as the state-of-the-art directed greybox fuzzer WAFLGo in 24 hours, even though WAFLGo started with a commit-reaching seed corpus in the majority of cases. If we amplify the \toolname-generated test cases using those as a seed corpus in coverage-guided greybox fuzzing, the number of bugs found doubles. We call the integration with fuzzing as \toolnamefuzz.

  In addition, we found that some commit messages are more expressive than others, thus we wonder how this impacts the effectiveness of \toolname. Our results above demonstrate that \toolname picks up on the change intention. For instance, if the commit message describes that this patch changes how floating point variables are treated in the \mujs JavaScript interpreter, then \toolname generates JavaScript programs that contain floating point variables. To study the impact of expressiveness, we change the commit messages minimally to reduce and increase the information in the commit message, respectively, and find a substantial impact on effectiveness. For instance, adding \review{17} words on average (max. \review{43}) to make ineffective commit messages more expressive \review{significantly increased} the number of bugs found.

\end{abstract}

\ccsdesc[500]{Software and its engineering~Software testing and debugging}
\ccsdesc[300]{Software and its engineering~Software evolution}

\keywords{Cleverest, Regression test generation, LLM}

 \maketitle

\section{Introduction}
Recently, Large Language Models (LLMs) have shown tremendous promise. Liu et al. \cite{liu2024largelanguagemodelbasedagents} published a comprehensive survey of research on the automation of software engineering processes, including requirements engineering, code generation, program analysis, testing, debugging, and end-to-end development and maintenance. %

In this paper, we explore the utility of LLMs as test generators for code commits or pull requests in the continuous integration/continuous development (CI/CD) pipeline.
When a new bug fix or a code change is committed and the project is built and analyzed as part of CI/CD pipeline, running an LLM for a few minutes to generate regression test cases for that commit could help exercise the changed code and potentially trigger any bugs.
Specifically, discovering bugs i) introduced or ii) insufficiently fixed by a commit are the two primary objectives of our paper. We investigate LLM-based regression test generation as a \emph{machine translation task} that takes the developer-provided commit message, the code change, and the name of a well-known input format (e.g., XML or JS) and produces regression test cases for the commit in the given input format.
In systems without precise test oracles (e.g., interpreters, parsers, compilers), such regressions are typically identified via crashes, sanitizer violations, or observable output differences, like in (differential) fuzzing. 

In our experiments, we tested \review{72} commits to \review{eight} programs (\mujs, \libxml, \poppler, \jerryscript, \zthree, \php, \review{\jq, and \micropython}), which take highly structured, human-readable input formats.
Our feedback-directed, zero-shot LLM-based prototype \toolname performed well. In under two minutes, on average, \toolname found as many bugs as the state-of-the-art directed greybox fuzzer WAFLGo found in 24 hours,
even though WAFLGo~\cite{Xiang:2024aa} already starts with a commit-reaching seed corpus for the majority of commits. We find that our tool's effectiveness varies across input formats and LLMs. For instance, \toolname performs much better for human-readable formats like XML and JavaScript than for binary formats like PDF. Effectiveness also scales with LLM size: a weaker \gptmini reduces effectiveness, while a stronger \deepseekr found a bug missed by \gpt.

Most interestingly, we identify a critical dependence of \toolname's effectiveness on  the commit message. In fact, \toolname continues to generate effective test cases even when only the commit message is provided, as long as the intention of the change is captured. For instance, if the commit message describes the handling of floating point variables, then \toolname generates Javascript programs that contain floating point variables. In contrast, the non-descriptive commit messages in \zthree (e.g., \textsf{fix \#[Issue ID]}) are unlikely to yield effective regression tests while the expressive commit messages are more likely to reveal bugs. To study the impact of expressiveness, we change the commit messages minimally to reduce and increase the information, respectively, and indeed confirm a substantial impact on effectiveness. For instance, adding \review{17} words on average (max. \review{43}) to make ineffective messages more expressive \review{significantly increased} the number of bugs found.

We also explore the marriage of greybox fuzzing and \toolname, called \toolnamefuzz, where the regression tests generated by \toolname are provided as initial seeds for subsequent greybox fuzzing. Indeed, if we amplify the \toolname-generated test cases in this way, we can find over twice the bugs as the directed fuzzer WAFLGo~\cite{Xiang:2024aa} with its default seed and \toolname alone.

\vspace{0.1cm}
\noindent
In summary, this paper makes the following contributions:
\begin{itemize}[leftmargin=*,itemsep=0.05cm]
  \item We introduce \toolname, a feedback-directed, zero-shot LLM-based regression test generation technique for programs that take highly structured, human-readable inputs to evaluate the utility of LLMs for regression test generation.
  \item We evaluate \toolname on \review{72} commits to eight popular programs and find that \toolname performs very well for programs using \icse{human-interpretable} file formats, even when only the commit message is given. This might mean, the LLM can translate the change \emph{intention} into a system input. We confirm this hypothesis by varying the expressiveness of the commit message.%
  \item \toolname performs comparably to WAFLGo, a SOTA few-shot regression test generator, while being substantially faster—even though WAFLGo's initial seeds are already close to bug-revealing. Using \toolname-generated tests as seeds for fuzzing finds more than twice as many bugs.

\end{itemize}

\begin{figure}[t]
  \includegraphics[width=.8\textwidth,clip,trim=10 10 10 10]{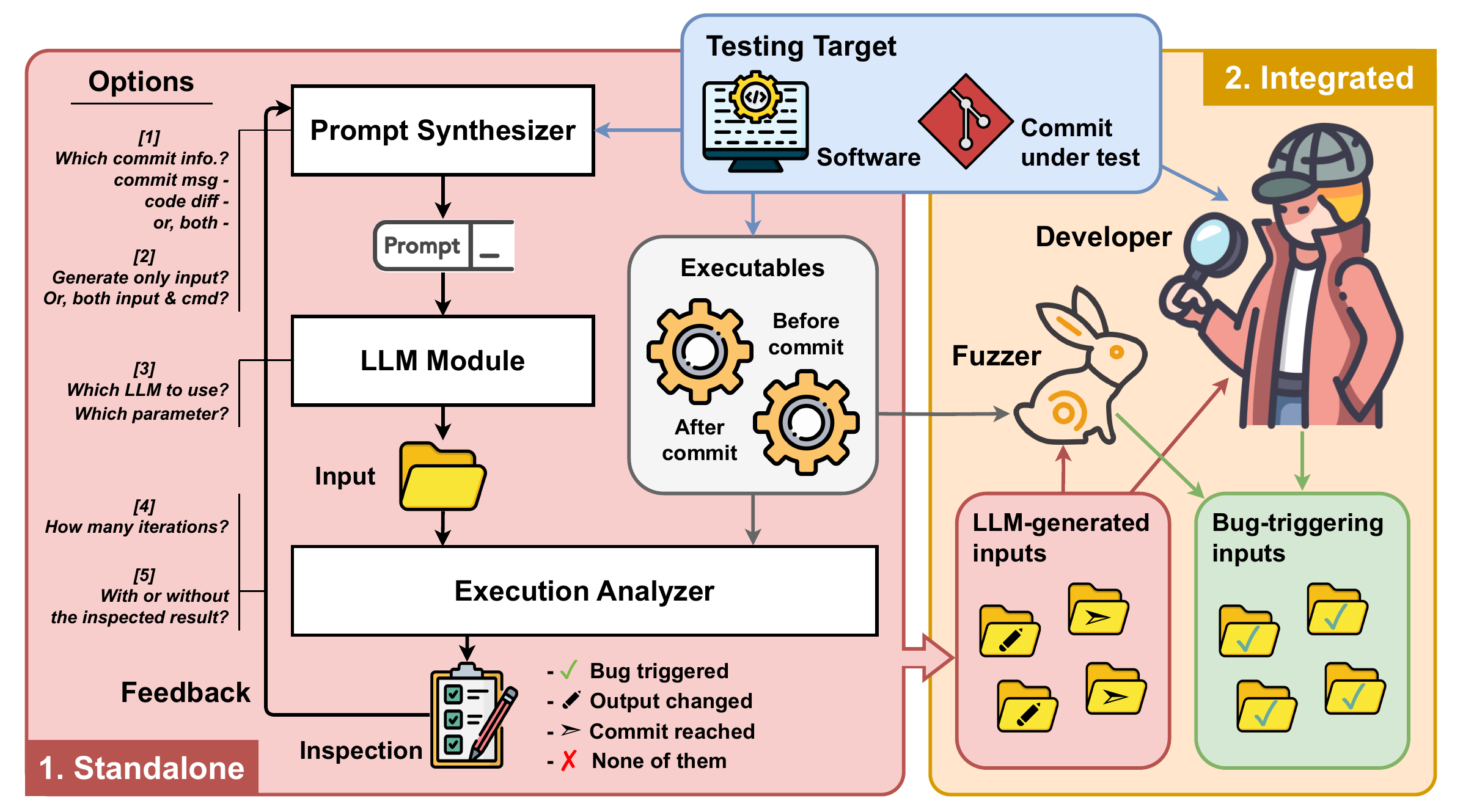}%
  \caption{\toolname: Our zero-shot feedback-guided LLM-based regression test generation methodology.}
  \label{fig:workflow}
\end{figure}

\section{Experimental Design}
In this paper, we seek to evaluate the capabilities of a large language model (LLM) as a regression test generation tool in the CI/CD pipeline for programs that take highly structured, human-readable inputs, like JavaScript programs or XML files. \review{While classical regression testing focuses on passing test cases across program versions, our setting lacks a specification or pre-existing tests. Thus, our primary objective is to determine how well an LLM performs in generating test cases that: reach the changed code, observe behavioral differences, and trigger bugs detectable via sanitizers. These metrics reflect standard practice in regression fuzzing, where a regression manifests as a newly introduced crash.}
We implement a feedback-directed LLM-based regression test generator involving a prompt synthesizer and execution analyzer. We call our tool \toolname. %

\subsection{Research Questions}
\begin{description}[leftmargin=0.4cm,itemsep=0.05cm]
  \item[RQ1.] \emph{How well does \toolname perform as a regression test generator?}
    We evaluate the two most important properties of \toolname if it was fully embedded in a CI/CD pipeline: effectiveness and execution time. Specifically, we evaluate how well \toolname can find bugs that were introduced in a commit (\emph{bug-finding}) and how well \toolname can reproduce bugs that were patched in a commit (\emph{bug-reproduction}).
    To understand how \toolname performs in the hands of a developer, we evaluate how ``close''
    \toolname-generated test cases are to revealing a bug—both quantitatively, using execution feedback distance, and qualitatively, through manual inspection.
  \item[{RQ2.}] \emph{How well does \toolname perform under various hyperparameter values? (Ablation Study)}
    Specifically, we analyze how well it performs compared to the default configuration if we
    (a)~only used the commit message but not the diff,
    (b)~only used the commit diff but not the message,
    (c)~let it generate the command in addition to the input,
    (d)~used maximum LLM temperature,
    (e)~used a less or a more powerful LLM (\gptmini vs \deepseekr),
    (f)~used 10 feedback loop iterations instead of 5, and
    (g)~dropped the execution analysis in the feedback loop.

  \item[{RQ3.}] \emph{How informative and self-contained do commit messages have to be to enable regression test generation without access to code?}
    Specifically, we investigate the impact of varying the descriptiveness of commit messages on \toolname's performance; our preliminary results from \textbf{RQ2} suggest that even with commit messages alone, \toolname can generate effective test cases. %

  \item[{RQ4.}] \emph{How does \toolname compare to WAFLGo, the state-of-the-art in regression test generation?}
    \review{WAFLGo \cite{Xiang:2024aa} is the state-of-the-art directed greybox fuzzer for regression test generation that was shown to outperform previous directed and regression greybox fuzzers \cite{aflchurn, aflgo, windranger, selectfuzz, fishfuzz}.}
    Specifically, we evaluate (1) the performance of WAFLGo as a few-shot regression test generator and (2) the dependence of WAFLGo on the initial seeds, and (3) the performance of \toolname against WAFLGo both directly and as a seed generator for zero-shot greybox fuzzing.
\end{description}
\textbf{More results}. Separately, we compare our LLM-based regression test generation approach with an \emph{undirected} LLM-based test generation approach, where no change information is provided—similar to existing methods explored in previous work~\cite{Fuzz4ALL,mokav,busybox}.
Moreover, we evaluate the performance of \toolname on the 50 most recent commits in \libxml. For both, we postpone the results to the \textbf{Supplementary Material} \cite{supplementary} for space reasons.

\subsection{\toolname: LLM-Based Regression Testing Technique and Implementation}

\autoref{fig:workflow} gives a procedural overview of \toolname, our zero-shot feedback-guided LLM-based regression test generation tool. It consists of three key components: \emph{Prompt Synthesizer}, \emph{LLM Module}, and \emph{Execution Analyzer} that work in a \emph{feedback loop} to generate test cases for a commit. %

\subsubsection{Prompt Synthesizer}
\label{sec:prompt-synthesizer}

\begin{figure}
  \begin{center}
    \includegraphics[width=.8\textwidth,clip,trim=20 10 20 40]{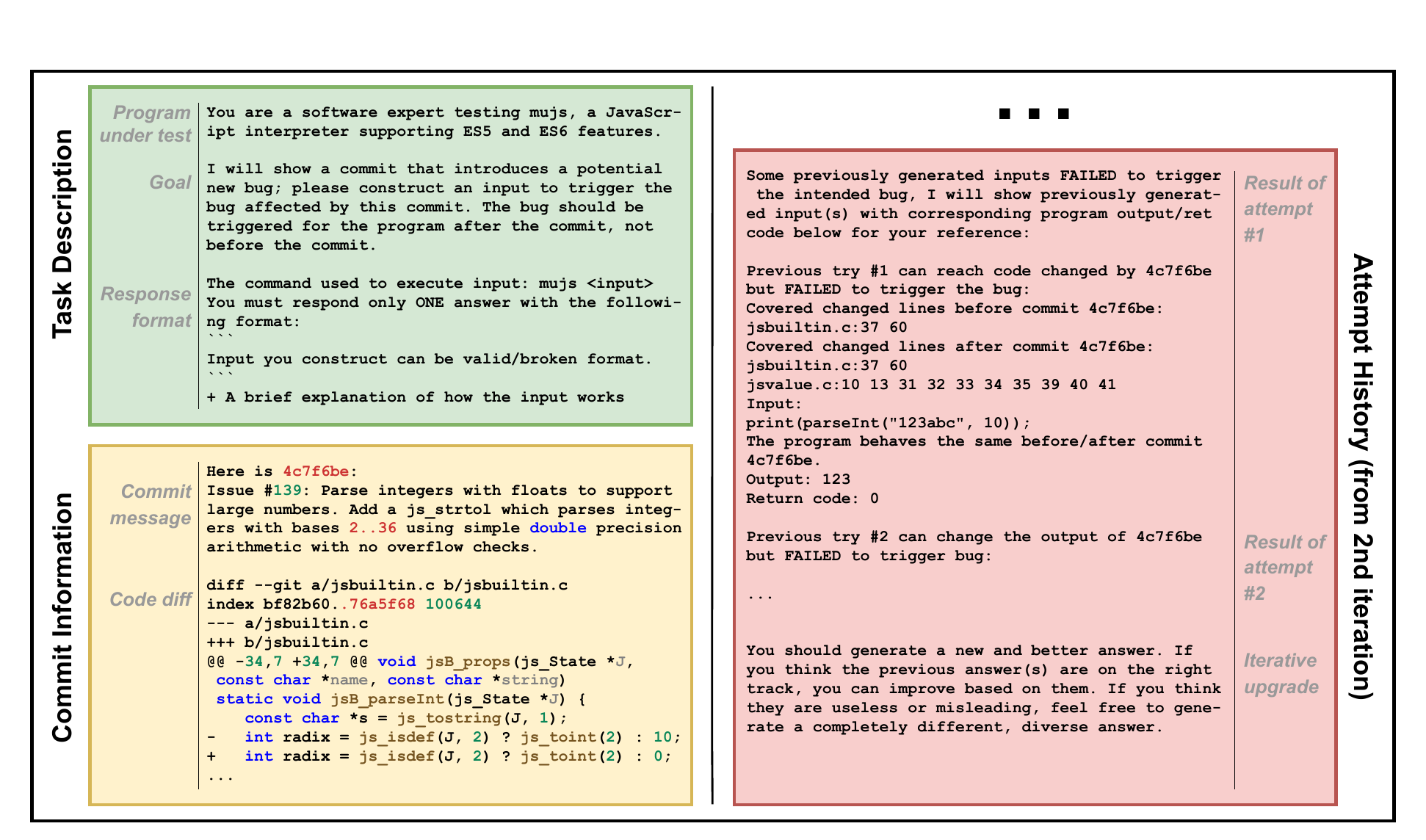}
    \caption{The prompt generated by the prompt synthesizer for the LLM to generate the test input for Mujs.}
    \label{fig:prompt0}
  \end{center}
\end{figure}

The \emph{Prompt Synthesizer} generates a domain-specific prompt for the LLM to create the test input. Given the testing target (i.e., the software and the commit under test) and optionally some feedback from a previous iteration, the prompt synthesizer constructs the prompt for the LLM, which includes the task description, commit information, and the attempt history.

Figure~\ref{fig:prompt0} shows an example prompt to generate a regression test case for a code commit to \mujs, a lightweight JavaScript interpreter. The synthesized prompt has the following structure:
\begin{enumerate}[leftmargin=0.55cm,itemsep=0.05cm,topsep=0.1cm]
  \item \textbf{Task Description}: The prompt starts with a general description of the task. The main parts consist of (a) a simple sentence to describe the program under test, e.g., ``JavaScript interpreter,'' (b) the goal of the LLM-generated input, e.g., ``trigger a bug introduced by the commit,'' and (c) the format specification of the LLM's response. %
  \item \textbf{Commit Information}: The second part contains the commit under test. We consider both the commit message and the code diff as part of the commit information, unless otherwise stated. The \emph{code diff} contains the changed lines plus a few lines of context. The \emph{commit message} offers a high-level description of the developer's intention and the purpose of the commit.
  \item \textbf{Attempt History}:
        When available, the prompt contains the execution result of the previously generated test case. Each item shows
        (a)~the general execution result from the execution analyzer (cf. Sec.~\ref{sec:exec-analzer}),
        (b)~the program output (incl. {\small\texttt{stdout}} \& {\small\texttt{stderr}}), and
        (c)~the return code.
\end{enumerate}
\textbf{Hyperparameters}. The prompt synthesizer can be configured differently affecting the task difficulty for the LLM. For instance, we may consider providing only the code diff, only the commit message, or both. We may provide the command-line utility and parameters ({\small\texttt{cmd}}) expected to be used to test the commit or let it generate itself: Notice that fuzzing can only modify the input and requires the user to specify the {\small\texttt{cmd}}. Yet, an LLM can generate the full {\small\texttt{cmd}} for the program execution. Allowing the generation of the {\small\texttt{cmd}} can make the task more challenging for the LLM, as it needs to understand the program's behavior to generate the correct {\small\texttt{cmd}}.

\subsubsection{LLM Module}

The generated prompt is fed into the LLM module, which generates the test input for the commit under test. The output of the LLM contains the content of the test input along with the explanation of the expected behavior and the {\small\texttt{cmd}} if asked. The output is then parsed to produce the actual test case, which is executed to verify the commit under test. Depending on the task description, the LLM tries to generate the input that triggers the bug introduced by the commit (without knowing a priori if the bug is introduced)
or reproduces the bug fixed by the commit.

The LLM module has several hyperparameters that affect the quality of the generated output. One is the \emph{size} (or capacity) of the LLM model: A larger model size generally results in better output quality but requires more computational resources. Another is the \emph{temperature}, which controls the randomness of the output: A higher temperature results in more randomness in the output.

\subsubsection{Execution Analyzer and Feedback Loop}
\label{sec:exec-analzer}

The \emph{Execution Analyzer} receives the LLM-generated test input and executes it on the program before and after the commit under test. The execution analyzer compares the two executions to measure if the generated input contributes to the commit testing. %
We measure the effectiveness using an ordinal scale:

\begin{itemize}[leftmargin=0.55cm,itemsep=0.05cm,topsep=0.1cm]
  \item[\cmark] \textbf{Bug Triggering}: To discover a bug if one exists is the ultimate objective of testing. Depending on the scenario, the execution analyzer checks if the input triggers the intended bug: it should trigger the bug after or before the commit in the bug-finding or bug-reproduction scenario, respectively, and should not trigger the same bug in the other program version. For instance, even if the input triggers a bug, for our intents and purposes it is not considered as bug triggering if the same bug is triggered in both versions of the program.
  \item[\dmark] \textbf{Output Changing}: Even if the input does not trigger a bug (e.g., if the sanitizer does not detect the anomaly), the input may witness a difference in behavior. We detect the behavior difference by comparing the output and return code of the program before and after the commit. Any difference will imply that the generated test successfully exposes the difference in the program behavior introduced by the commit.
  \item[\rmark] \textbf{Commit Reaching}: The first necessary condition for an input to reveal a bug introduced by the commit is to reach the code changed by the commit. Yet, even generating an input that reaches the commit is not trivial. We thus measure the coverage of the program execution with the generated input to see if there is any overlap between the code changed by the commit and the code covered by the execution.
\end{itemize}
The output of the execution analyzer is the assessment result of the generated input categorized into four types: bug triggered, output changed, commit reached, and none of the above. Notice that the prior categories are strict inclusions of the latter ones.

If the execution results do not trigger a bug, the inspected execution result (i.e., which lines of the commit the input covers if at all, and whether the output is different) is provided to the prompt synthesizer for the next input generation. This incremental prompting strategy guides the LLM in exploring the input space more effectively and allows it to generate high-quality test input for the commit under test. \review{This is also confirmed by recent works~\cite{panta, mutgen, MuTAP} that showed LLM-based test generation benefits substantially from feedback loops that augment prompts with analysis results. 
}

Two hyperparameters in the feedback loop affect the performance of \toolname. One hyperparameter allows to enable or disable the feedback loop. If disabled, only the previously generated input is used as the prompt for the next iteration to avoid generating the same input. The other hyperparameter is the number of iterations, which determines the number of times the LLM generates the input, and the execution analyzer verifies the input. A higher number of iterations can lead to more effective input generation but requires more computational resources.

\subsubsection{Implementation}

For \emph{commit information}, we use Git to extract the commit message and code diff. The command `{\small\texttt{git show  --format=\%B}}' is used to print them together in a concise format. We remove all changes that are not related to the source code.

To \emph{detect whether a bug is triggered}, our approach relies on an automated failure oracle that flags failures when executing generated test cases. \review{Any mechanism that can automatically identify failures—such as sanitizers, assertions, or runtime checks—can be integrated without changes to the test generation pipeline.} In this study, we use AddressSanitizer (ASan) \cite{asan}, which inserts instrumentation during compilation to detect memory-corruption at runtime, including buffer overflows and use-after-free errors, without requiring manually written oracles.
To \emph{detect behavior differences}, we compare the {\small\texttt{stdout}}, {\small\texttt{stderr}}, and return code of the program built before and after the commit under test.
To \emph{detect whether the changed code is reached}, we use {\small\texttt{GCOV}} to get code coverage information. We consider the input reaches the commit if there is any overlap between the code changed by the commit and the code covered by the execution.

For our \emph{LLM interactions}, we use {\small\texttt{openai-cli}}, a command-line client. We use \gpt as the default model and set the max token to 4096 to achieve a balance between performance and cost. Notably, we do not keep conversation history when querying LLM, as it consumes more tokens. The default temperature is set to 0.5, and the number of iterations is set to 5. We additionally consider up to ten iterations and the temperature to be 1.0 to examine parameter effects. We also consider two other models, \gptmini and \deepseekr, to evaluate the impact of model size on performance. \gptmini, a smaller and more efficient version of \gpt,
which is designed especially for applications where affordability and lower latency are critical.
\deepseekr, recently released (Jan. 2025), is the cutting-edge LLM model especially designed for reasoning tasks. It has 671B parameters and scores 90.8\% on the MMLU benchmark. \review{As a reasoning model, \deepseekr returns both the final model output and intermediate reasoning traces, but, since they are returned separately, we exclusively retrieved the final generated test input and ignored the intermediate reasoning output. This strategy allowed us to prevent excessively long reasoning chains from affecting the correctness or structure of the generated regression test cases.}

\subsection{Benchmark Selection and Infrastructure}
\label{sec:benchmarks}

To answer our research questions, we choose a commit benchmark dataset for our experiment based on the following \emph{selection criteria}:
\begin{enumerate}[leftmargin=0.55cm,itemsep=0.05cm,topsep=0.1cm]
  \item We want to focus on regression test generation for programs that take highly structured, human-readable input formats.
  \item We want to maximize the fairness of our comparison to the state-of-the-art regression test generation tool WAFLGo, which is used for comparison.
  \item We consider real-world regression bugs of various types that have been introduced and patched in identified commits to open-source software.
\end{enumerate}

\begin{table}[t]
  \centering
  \caption{Detail of Bug Dataset.
    The three last columns show the ID in the issue tracker and the commit IDs of the bug-introducing-commit (BIC) and the bug-fixing-commit (BFC).}
  \label{tab:dataset}
  \begin{subtable}[b]{.525\linewidth}
    \centering
    \resizebox{\textwidth}{!}{%
      \begin{tabular}{@{}c|c|l|c|c|c@{}}
        \toprule
        \textbf{Software}         & \textbf{Description}                                                             & \multicolumn{1}{c|}{\textbf{Command-Line Utility and Parameters}}                                                                                                & \textbf{Issue}         & \textbf{BIC}             & \textbf{BFC}             \\ \midrule
        \multirow{4}{*}{\mujs}    & \multirow{4}{*}{\begin{tabular}[c]{@{}c@{}}JavaScript\\interpreter\end{tabular}} & \texttt{mujs} $\langle$input file$\rangle$                                                                                                                       & \#65                   & 8c27b12                  & 833f82c                  \\
                                  &                                                                                  & \texttt{mujs} $\langle$input file$\rangle$                                                                                                                       & \#141                  & 832e069                  & 6871e5b                  \\
                                  &                                                                                  & \texttt{mujs} $\langle$input file$\rangle$                                                                                                                       & \#145                  & 4c7f6be                  & f93d245                  \\
                                  &                                                                                  & \texttt{mujs} $\langle$input file$\rangle$                                                                                                                       & \#166                  & 3f71a1c                  & 8b5ba20                  \\ \midrule
        \multirow{6}{*}{\libxml}  & \multirow{6}{*}{\begin{tabular}[c]{@{}c@{}}XML\\parsing\\library\end{tabular}}   & \texttt{xmllint --recover --sax1 --sax} $\langle$input file$\rangle$                                                                                             & \#535                  & 9a82b94                  & d0c3f01                  \\
                                  &                                                                                  & \multirow{2}{*}{\begin{tabular}[c]{@{}c@{}}\texttt{xmllint --recover --dropdtd --sax1}\\\texttt{--nofixup-base-uris} $\langle$input file$\rangle$ \end{tabular}} & \multirow{2}{*}{\#550} & \multirow{2}{*}{74c84a8} & \multirow{2}{*}{6273df6} \\
                                  &                                                                                  &                                                                                                                                                                  &                        &                          &                          \\
                                  &                                                                                  & \texttt{xmllint --encode UTF-16 --maxmem 325 --encode UTF-16} $\langle$input file$\rangle$                                                                       & \#553                  & 1406b20                  & db21cd5                  \\
                                  &                                                                                  & \texttt{xmllint --htmlout} $\langle$input file$\rangle$                                                                                                          & \#720                  & 14604a4                  & 3ad7f81                  \\
                                  &                                                                                  & \texttt{xmllint -repeat -push} $\langle$input file$\rangle$                                                                                                      & \#841                  & 9cfc723                  & dc6270d                  \\ \midrule
        \multirow{5}{*}{\poppler} & \multirow{5}{*}{\begin{tabular}[c]{@{}c@{}}PDF\\rendering\\library\end{tabular}} & \texttt{pdfunite} $\langle$input file 1$\rangle$ $\langle$input file 2$\rangle$ \texttt{output.pdf}                                                              & \#1282                 & 3d35d20                  & 4564a00                  \\
                                  &                                                                                  & \texttt{pdfunite} $\langle$input file 1$\rangle$ $\langle$input file 2$\rangle$ \texttt{output.pdf}                                                              & \#1289                 & 3cae777                  & efb6868                  \\
                                  &                                                                                  & \texttt{pdftops} $\langle$input file$\rangle$ \texttt{/dev/null}                                                                                                 & \#1303                 & e674ca6                  & a4ca3a9                  \\
                                  &                                                                                  & \texttt{pdftoppm -mono -cropbox} $\langle$input file$\rangle$                                                                                                    & \#1305                 & aaf2e80                  & 907d05a                  \\
                                  &                                                                                  & \texttt{pdftoppm -mono -cropbox} $\langle$input file$\rangle$                                                                                                    & \#1381                 & 245abad                  & 1be35ee                  \\ \midrule
        \multirow{4}{*}{\jerryscript} & \multirow{4}{*}{\begin{tabular}[c]{@{}c@{}}JavaScript\\interpreter\end{tabular}} & \texttt{jerry} $\langle$input file$\rangle$                       & \#5013         & b7e3bae      & 3686410      \\
                                      &                                                                                  & \texttt{jerry} $\langle$input file$\rangle$                       & \#5138         & 70e275e      & 95cc5e9      \\
                                      &                                                                                  & \texttt{jerry} $\langle$input file$\rangle$                       & \#5117         & 53b61c1      & d7e2125      \\
                                      &                                                                                  & \texttt{jerry} $\langle$input file$\rangle$                       & \#5153         & 70e275e      & de51531      \\ \bottomrule
      \end{tabular}}
  \end{subtable}
  \begin{subtable}[b]{.46\linewidth}
    \centering
    \resizebox{\textwidth}{!}{%
      \begin{tabular}{@{}c|c|l|c|c|c@{}}
        \toprule
        \textbf{Software}             & \textbf{Description}                                                             & \multicolumn{1}{c|}{\textbf{Command-Line Utility and Parameters}} & \textbf{Issue} & \textbf{BIC} & \textbf{BFC} \\ \midrule
        \multirow{4}{*}{\zthree}      & \multirow{4}{*}{\begin{tabular}[c]{@{}c@{}}Theorem\\prover\end{tabular}}         & \texttt{z3} $\langle$input file$\rangle$                          & \#6659         & 0f3c562      & a849a29      \\
                                      &                                                                                  & \texttt{z3} $\langle$input file$\rangle$                          & \#6914         & a62e4b2      & eea9c0b      \\
                                      &                                                                                  & \texttt{z3} $\langle$input file$\rangle$                          & \#7246         & dec87fe      & 49610f5      \\
                                      &                                                                                  & \texttt{z3} $\langle$input file$\rangle$                          & \#7252         & a7b1db6      & a6b5027      \\ \midrule
        \multirow{4}{*}{\php}         & \multirow{4}{*}{\begin{tabular}[c]{@{}c@{}}PHP\\interpreter\end{tabular}}        & \texttt{php} $\langle$input file$\rangle$                         & \#16777        & 5cb38e9      & 18b18f0      \\
                                      &                                                                                  & \texttt{php} $\langle$input file$\rangle$                         & \#16978        & 8c704ab      & d17ed34      \\
                                      &                                                                                  & \texttt{php} $\langle$input file$\rangle$                         & \#17442        & 998bce1      & b068c2f      \\
                                      &                                                                                  & \texttt{php} $\langle$input file$\rangle$                         & \#17463        & 7904a08      & e4473ab      \\ \midrule
        \multirow{4}{*}{\jq}      & \multirow{4}{*}{\begin{tabular}[c]{@{}c@{}}Command-line\\JSON\\processor\end{tabular}} & \texttt{jq -nf} $\langle$input file$\rangle$                                                                                                                     & \#2825                 & 680baef                  & 3fa10e8                  \\
                                  &                                                                                  & \texttt{jq .} $\langle$input file$\rangle$                                                                                                                       & \#2976                 & 680baef                  & 71c2ab5                  \\
                                  &                                                                                  & \texttt{jq .[54E100]=7} $\langle$input file$\rangle$                                                                                                             & \#3262                 & 0e067ef                  & de21386                  \\
                                  &                                                                                  & \texttt{jq -nf} $\langle$input file$\rangle$                                                                                                                     & \#49014\footnotemark & 4003202                  & 499c91b                  \\ \midrule
        \multirow{6}{*}{\micropython} & \multirow{6}{*}{\begin{tabular}[c]{@{}c@{}}Python for\\microcontrollers\end{tabular}} & \texttt{micropython} $\langle$input file$\rangle$                 & \#13007        & 4c93955      & f397a3e      \\
                                      &                                                                                  & \texttt{micropython} $\langle$input file$\rangle$                 & \#13041        & 1fef566      & 908ab1c      \\
                                      &                                                                                  & \texttt{micropython} $\langle$input file$\rangle$                 & \#17733        & 5e9189d      & 3185bb5      \\
                                      &                                                                                  & \texttt{micropython} $\langle$input file$\rangle$                 & \#17815        & 69ead7d      & c0252d7      \\
                                      &                                                                                  & \texttt{micropython} $\langle$input file$\rangle$                 & \#17841        & e702046      & 4b013ec      \\
                                      &                                                                                  & \texttt{micropython} $\langle$input file$\rangle$                 & \#17847        & 7e14680      & 0615d13      \\ \bottomrule
      \end{tabular}}
  \end{subtable}
\end{table}
\footnotetext{\jq~\#49014 refers to CVE-2025-49014, which is publicly available on GitHub security advisories but not issue tracker.}

\noindent
\autoref{tab:dataset} details the dataset that meets our selection criteria. Specifically, we found that the benchmark dataset used in WAFLGo's evaluation~\cite{Xiang:2024aa} satisfies criteria (2) and (3), while three programs (the JavaScript interpreter \mujs, the XML parsing library \libxml, and the PDF rendering library \poppler) satisfy criterion (1).
\icse{Additionally, we added \review{five} more subject programs—the JavaScript interpreter \jerryscript, the SMT solver \zthree, the PHP interpreter \php, \review{the JSON processor \jq, and the Python interpreter \micropython}—which are frequently used in software testing papers \cite{gramatron,Fuzz4ALL}.}
These programs cover various input formats. Each software has one or more command-line utilities that retrieve textual input files as command-line parameters with relevant options.

The WAFLGo dataset lists 11 memory-related bugs for these three benchmark programs. \icse{For the \review{five} newly added programs, we searched GitHub issues and selected the \review{four to five} newest reproducible bugs for each.} \review{We also added three extra bugs for \libxml to make it also have five bugs.} The bug types span a range of memory-related issues, including buffer overflows and use-after-frees.
For each bug, \review{we intentionally include both bug-introducing commits (BICs) and bug-fixing commits (BFCs). BICs are used to evaluate the \emph{bug-finding} scenario, modeling realistic feature or refactoring changes that may introduce regressions. BFCs are used to evaluate the \emph{bug-reproduction} scenario, representing cases where no regression test existed prior to the fix; the objective is to automatically synthesize a regression test that would prevent re-introduction of the fixed bug.} In total, we have \review{72} commits in the experiment.

\subsubsection*{Experiment Configuration}
The default configuration of our \toolname is as follows: We provide the LLM with both the commit message and code diff as commit information, along with the command-line utility and parameters used to test the commit. Execution analysis results are included in feedback. \review{We use the \gpt model with 0.5 temperature to balance creativity and precision, following ChatAFL~\cite{chatafl}. We limit each run to a safe upper bound of five feedback iterations based on our preliminary study, where we observed that effective test cases were generated within 1.67 iterations on average.}
Each LLM-based input generation has a 30-second timeout.\footnote{\review{Due to free-tier API limitations, DeepSeek-R1 experiments had a median execution time of approximately 15 minutes; commercial deployments would substantially reduce this latency.}}
To account for randomness in the LLM and fuzzer, we repeat each experiment ten times.
All experiments run in a Docker container with 64 AMD EPYC 7713P cores @ 2.0GHz and 251GB memory.

\section{Experimental Results}
\subsection{RQ1. Evaluation of Capabilities}
\label{sec:rq1}

\begin{table}
  \caption{Results for bug finding and bug reproduction effectiveness of \toolname across ten repetitions. For each bug and scenario, we report the average effectiveness score on a slider (\xmark: Not reached; \rmark: Reached; \dmark: Output-changing; \cmark: Bug-revealing), the number of trials where the commit was reached (Reach), output was changed (Change), bug was found (Bug), and the average time in seconds (T (s)).}
  \label{tab:rq1}
  \vspace{-0.5em}
  \resizebox{.95\textwidth}{!}{%
    \begin{tabular}{cc|lrrrr|lrrrr}
      \toprule
                                            &                              & \multicolumn{5}{c|}{\textbf{Bug-Introducing-Commit}} & \multicolumn{5}{c}{\textbf{Bug-Fixing Commit}}                                                                               \\
                                            &                              & \xmark~\rmark~\dmark~\cmark                       & Reach & Change  & Bug                                            & T (s)                        & \xmark~\rmark~\dmark~\cmark   & Reach & Change & Bug   & T (s) \\
      \midrule
      \multirow{4}{*}{\mujs} & \#65 & \SLIDER{3.3em}{1.00}{circle} & 10/10 & 10/10 & 10/10 & 7.5 & \SLIDER{3.3em}{0.73}{circle} & 10/10 & 6/10 & 6/10 & 19.1 \\
      & \#141 & \SLIDER{3.3em}{0.67}{circle} & 10/10 & 10/10 & 0/10 & 45.4 & \SLIDER{3.3em}{0.70}{circle} & 10/10 & 10/10 & 1/10 & 31.7 \\
      & \#145 & \SLIDER{3.3em}{0.87}{circle} & 10/10 & 8/10 & 8/10 & 37.1 & \SLIDER{3.3em}{0.93}{circle} & 10/10 & 9/10 & 9/10 & 28.8 \\
      & \#166 & \SLIDER{3.3em}{0.37}{circle} & 10/10 & 1/10 & 0/10 & 57.6 & \SLIDER{3.3em}{1.00}{circle} & 10/10 & 10/10 & 10/10 & 7.4 \\
      \midrule
      \multirow{5}{*}{\libxml} & \#535 & \SLIDER{3.3em}{1.00}{circle} & 10/10 & 10/10 & 10/10 & 5.4 & \SLIDER{3.3em}{1.00}{circle} & 10/10 & 10/10 & 10/10 & 5.5 \\
      & \#550 & \SLIDER{3.3em}{0.40}{circle} & 10/10 & 2/10 & 0/10 & 27.6 & \SLIDER{3.3em}{0.00}{circle} & 0/10 & 0/10 & 0/10 & 29.9 \\
      & \#553 & \SLIDER{3.3em}{1.00}{circle} & 10/10 & 10/10 & 10/10 & 6.7 & \SLIDER{3.3em}{1.00}{circle} & 10/10 & 10/10 & 10/10 & 13.7 \\
      & \#720 & \SLIDER{3.3em}{0.33}{circle} & 10/10 & 0/10 & 0/10 & 53.6 & \SLIDER{3.3em}{0.23}{circle} & 7/10 & 0/10 & 0/10 & 104.1 \\
      & \#841 & \SLIDER{3.3em}{1.00}{circle} & 10/10 & 10/10 & 10/10 & 5.9 & \SLIDER{3.3em}{1.00}{circle} & 10/10 & 10/10 & 10/10 & 5.5 \\
      \midrule
      \multirow{5}{*}{\poppler} & \#1282 & \SLIDER{3.3em}{0.37}{circle} & 8/10 & 3/10 & 0/10 & 196.6 & \SLIDER{3.3em}{0.30}{circle} & 9/10 & 0/10 & 0/10 & 123.5 \\
      & \#1289 & \SLIDER{3.3em}{0.13}{circle} & 4/10 & 0/10 & 0/10 & 165.0 & \SLIDER{3.3em}{0.67}{circle} & 8/10 & 6/10 & 6/10 & 90.8 \\
      & \#1303 & \SLIDER{3.3em}{0.00}{circle} & 0/10 & 0/10 & 0/10 & 218.7 & \SLIDER{3.3em}{0.07}{circle} & 1/10 & 1/10 & 0/10 & 132.6 \\
      & \#1305 & \SLIDER{3.3em}{0.00}{circle} & 0/10 & 0/10 & 0/10 & 267.2 & \SLIDER{3.3em}{0.00}{circle} & 0/10 & 0/10 & 0/10 & 147.7 \\
      & \#1381 & \SLIDER{3.3em}{0.00}{circle} & 0/10 & 0/10 & 0/10 & 202.8 & \SLIDER{3.3em}{0.00}{circle} & 0/10 & 0/10 & 0/10 & 147.1 \\
      \midrule
      \multirow{4}{*}{\jerryscript} & \#5013 & \SLIDER{3.3em}{0.33}{circle} & 10/10 & 0/10 & 0/10 & 106.8 & \SLIDER{3.3em}{0.33}{circle} & 10/10 & 0/10 & 0/10 & 72.0 \\
      & \#5117 & \SLIDER{3.3em}{0.67}{circle} & 10/10 & 10/10 & 0/10 & 71.4 & \SLIDER{3.3em}{0.67}{circle} & 10/10 & 5/10 & 5/10 & 50.8 \\
      & \#5138 & \SLIDER{3.3em}{0.73}{circle} & 10/10 & 10/10 & 2/10 & 78.7 & \SLIDER{3.3em}{0.33}{circle} & 10/10 & 0/10 & 0/10 & 74.2 \\
      & \#5153 & \SLIDER{3.3em}{0.67}{circle} & 10/10 & 10/10 & 0/10 & 95.2 & \SLIDER{3.3em}{0.00}{circle} & 0/10 & 0/10 & 0/10 & 39.4 \\
      \midrule
      \multirow{4}{*}{\zthree} & \#6659 & \SLIDER{3.3em}{0.00}{circle} & 0/10 & 0/10 & 0/10 & 136.2 & \SLIDER{3.3em}{0.00}{circle} & 0/10 & 0/10 & 0/10 & 74.7 \\
      & \#6914 & \SLIDER{3.3em}{0.07}{circle} & 2/10 & 0/10 & 0/10 & 188.9 & \SLIDER{3.3em}{0.00}{circle} & 0/10 & 0/10 & 0/10 & 76.3 \\
      & \#7246 & \SLIDER{3.3em}{0.33}{circle} & 10/10 & 0/10 & 0/10 & 75.6 & \SLIDER{3.3em}{0.00}{circle} & 0/10 & 0/10 & 0/10 & 68.6 \\
      & \#7252 & \SLIDER{3.3em}{0.27}{circle} & 8/10 & 0/10 & 0/10 & 100.6 & \SLIDER{3.3em}{0.17}{circle} & 5/10 & 0/10 & 0/10 & 76.1 \\
      \midrule
      \multirow{4}{*}{\php} & \#16777 & \SLIDER{3.3em}{0.33}{circle} & 10/10 & 0/10 & 0/10 & 109.9 & \SLIDER{3.3em}{0.33}{circle} & 10/10 & 0/10 & 0/10 & 87.7 \\
      & \#16978 & \SLIDER{3.3em}{0.37}{circle} & 10/10 & 1/10 & 0/10 & 106.0 & \SLIDER{3.3em}{0.00}{circle} & 0/10 & 0/10 & 0/10 & 70.7 \\
      & \#17442 & \SLIDER{3.3em}{0.67}{circle} & 10/10 & 10/10 & 0/10 & 103.3 & \SLIDER{3.3em}{0.00}{circle} & 0/10 & 0/10 & 0/10 & 78.4 \\
      & \#17463 & \SLIDER{3.3em}{0.20}{circle} & 6/10 & 0/10 & 0/10 & 122.7 & \SLIDER{3.3em}{1.00}{circle} & 10/10 & 10/10 & 10/10 & 9.1 \\
      \midrule
      \multirow{4}{*}{\jq} & \#2825 & \SLIDER{3.3em}{0.33}{circle} & 10/10 & 0/10 & 0/10 & 23.6 & \SLIDER{3.3em}{0.37}{circle} & 10/10 & 1/10 & 0/10 & 28.4 \\
      & \#2976 & \SLIDER{3.3em}{0.33}{circle} & 10/10 & 0/10 & 0/10 & 22.3 & \SLIDER{3.3em}{1.00}{circle} & 10/10 & 10/10 & 10/10 & 3.8 \\
      & \#3262 & \SLIDER{3.3em}{1.00}{circle} & 10/10 & 10/10 & 10/10 & 9.7 & \SLIDER{3.3em}{1.00}{circle} & 10/10 & 10/10 & 10/10 & 8.4 \\
      & \#49014 & \SLIDER{3.3em}{0.13}{circle} & 2/10 & 2/10 & 0/10 & 32.2 & \SLIDER{3.3em}{0.00}{circle} & 0/10 & 0/10 & 0/10 & 32.7 \\
      \midrule
      \multirow{6}{*}{\micropython} & \#13007 & \SLIDER{3.3em}{0.53}{circle} & 9/10 & 7/10 & 0/10 & 27.0 & \SLIDER{3.3em}{0.07}{circle} & 1/10 & 1/10 & 0/10 & 26.6 \\
      & \#13041 & \SLIDER{3.3em}{0.00}{circle} & 0/10 & 0/10 & 0/10 & 31.0 & \SLIDER{3.3em}{0.67}{circle} & 10/10 & 10/10 & 0/10 & 21.8 \\
      & \#17733 & \SLIDER{3.3em}{0.33}{circle} & 10/10 & 0/10 & 0/10 & 26.3 & \SLIDER{3.3em}{0.33}{circle} & 10/10 & 0/10 & 0/10 & 31.4 \\
      & \#17815 & \SLIDER{3.3em}{0.63}{circle} & 10/10 & 9/10 & 0/10 & 26.3 & \SLIDER{3.3em}{1.00}{circle} & 10/10 & 10/10 & 10/10 & 5.5 \\
      & \#17841 & \SLIDER{3.3em}{0.67}{circle} & 10/10 & 10/10 & 0/10 & 21.4 & \SLIDER{3.3em}{0.33}{circle} & 10/10 & 0/10 & 0/10 & 39.2 \\
      & \#17847 & \SLIDER{3.3em}{0.67}{circle} & 10/10 & 10/10 & 0/10 & 40.9 & \SLIDER{3.3em}{0.00}{circle} & 0/10 & 0/10 & 0/10 & 29.2 \\
      \midrule
      \multicolumn{2}{c|}{\textbf{Average}} & \SLIDER{3.3em}{0.46}{circle} & \textbf{31/36} & \textbf{20/36} & \textbf{7/36} & \textbf{79.3} & \SLIDER{3.3em}{0.42}{circle} & \textbf{25/36} & \textbf{17/36} & \textbf{13/36} & \textbf{52.6} \\
      \bottomrule
    \end{tabular}}
    \vspace{-1em}
\end{table}

\textbf{Effectiveness}.
We measure the effectiveness of \toolname by computing the \emph{average effectiveness score} across all ten repetitions. If the generated regression test case finds the bug in the bug-introducing commit (BIC) or reproduces the bug in the bug-fixing commit (BFC), the score is 3 (\cmarknospace). Similarly, 2 indicates an output-changing test case (\dmarknospace), 1 indicates a commit-reaching test case (\rmarknospace), and 0 indicates a test case that fails to reach the changed code (\xmarknospace). Visually, we represent the average effectiveness score on a slider. The slider is colored in {\color{red}red}, {\color{orange}orange}, {\color{olive}olive}, and {\color{teal}teal} if the score is within the range of [0-0], (0-1], (1-2], and (2-3], respectively.

\autoref{tab:rq1} shows how well \toolname performs as a regression test generator.
We find that \toolname performs well in bug finding and bug reproduction for \icse{\review{four programs}—\mujs, \libxml, \jerryscript, \review{and \jq—}that use more \icse{human-interpretable} formats}. Despite lacking access to the full program code and without explicit information about the input features needed to exercise the changed code, \toolname identified bugs in \review{7} out of \review{17} BICs and reproduced the bugs patched in \review{10} out of \review{17} BFCs. Furthermore, in \review{14} out of the remaining \review{17} commits of \review{these four programs}, it at least reached the changed code. It also performs meaningfully well for \php and \review{\micropython, reaching the changed code in \review{16} out of \review{20} commits and reproducing bugs in 2 BFCs.} \review{\toolname shows consistent high reachability for these two interpreter programs, yet bug-triggering can be harder due to their complex language features. For example, in PHP \#16777, \toolname successfully reached the vulnerable logic by re-invoking constructor on attached \texttt{DOMElement}, but the UAF bug can only manifest with specific initialization order.}
Note that \toolname performs system-level testing, hence, no false positives in our result. We also confirmed that there is no flakiness.

However, for \poppler PDF parser, which requires a complex input format, \toolname could only reproduce one bug. At least, it reached the changed code in half of the commits and exposed a difference in one commit. \toolname consistently fails to generate an effective input in both scenarios for issues \#1305 and \#1381 of \poppler.
While the generated input contains some related components required to test the commit, it does not satisfy the validity constraints required to successfully parse the input.
For instance, \toolname can generate an input that already contains the key elements necessary for revealing the bug in Issue \#1305, including an annotation of type {\small\texttt{Highlight}} with an appearance stream and a {\small\texttt{Resources}} dictionary with an {\small\texttt{ExtGState}} entry. The failure to reach the commit occurs because Poppler strictly checks the existence of {\small\texttt{QuadPoints}} and {\small\texttt{Rect}} properties in the annotation object and exits early before reaching the changed code. If \toolname knew these validity constraints and added these specific elements, as we confirmed, the generated PDF input could have reached the commit and even demonstrated an output difference.

\toolname also struggles to generate bug-reproducing inputs for \zthree due to extremely simple commit messages that provide no information about the bug or fix. All four commits use the generic message ``fix \#[Issue ID],'' with no further details. These are clear outliers (word count: 2) compared to the average commit message length in other programs (word count: \review{38}).\footnote{Full commit message length statistics are available in Supplementary Material \cite{supplementary}.}
This is another piece of evidence for our observation that the LLM can translate the \emph{intention} of a well-described code change into a regression test case.
We investigate this phenomenon further in \textbf{RQ3}.

\textbf{Execution time and cost}.
\toolname is well-suited for the time-constrained environment of a CI/CD pipeline. \review{Execution times reported below include the full pipeline (commit extraction, test generation, and coverage collection).} On average, it takes \icse{mostly} less than one minute to generate XML, JavaScript, \review{JSON, and Python} inputs for \mujs, \libxml, \jerryscript, \review{\jq, and \micropython} ($\sim$$0.2\$$ on OpenAI) and still less than five minutes for PDF, PHP, and SMTLIB2 inputs, which involve different structural requirements ($\sim$$0.5\$$).

\textbf{Utility of \toolname-generated test cases}.
We evaluate how developers can use the \toolname-generated input in the pursuit of commit testing. %
Based on our execution analysis result, we found that even when \toolname-generated inputs do not directly trigger the bug, the commit-reaching or output-changing inputs are still useful for the human developer to understand the program behavior and to guide the bug-finding process. \toolname-generated test cases for BICs often ``prepare'' the precondition for the input to test the commit. For instance, the commit introducing Issue~\#550 of \libxml refines the XML parser's logic for checking the occurrence of the `<' character in entities. \toolname catches the \textit{``intention'' of the commit} and generates an input containing a `<' character for the entity (e.g., \toolname generated {\small\texttt{<!ENTITY test "<notallowed>">}}) to reach the commit.
The bug appears with the command {\small\texttt{xmllint --dropdtd}}, which removes the Document Type Definition (DTD) section from the document, which includes the entity definitions; thus, if \toolname-generated input has an entity reference, which gets removed by the command (e.g., {\small\texttt{<!DOCTYPE[<!ENTITY test>]<o><t test=``\&test;''>...}}), the bug would have been exposed.

\toolname-generated test cases for BFCs can often be modified to trigger the bug. For instance, the commit that fixes Issue~\#141 of \mujs adds missing end-of-string checks in the regexp lexer for special syntax-indicating starting characters, including $\backslash$x, $\backslash$u, and $\backslash$c. \toolname generates the input {\small\texttt{var regex = /$\backslash$x/;}} which exactly reflects the changed feature and produces a difference in program output from `\textsf{SyntaxError: invalid escape sequence}` to `\textsf{SyntaxError: unterminated escape sequence},` demonstrating a newly added error-handling sequence in the commit. To trigger the bug in the program before the BFC, the regex has to have a very long string that starts with one of those special characters---information that cannot be derived from the commit information alone. Yet, the human developer can easily modify the \toolname-generated input to contain a long string that reproduces the bug that was fixed by that commit. We further investigate the utility of \toolname-generated test cases in \textbf{RQ4} as fuzzing seeds.

\result{\textbf{RQ1}. \toolname performs well for the commits to programs that take more \icse{human-interpretable} formats but struggled to generate the right structure for the more complex format. With a short execution time of minutes, \toolname is well-suited to be used in the CI/CD pipeline. As they are easy to read and may already be halfway there, \toolname-generated inputs can also be used as a starting point for manual regression testing, even if they do not trigger any bugs in the given commit.}

\subsection{RQ2. Ablation Study}

\autoref{tab:rq2} presents the results of an ablation study on how \toolname performs under various configurations and hyperparameter values for the bug-finding and bug-reproduction scenarios. Specifically, we evaluate how well \toolname performs compared to the default configuration if we
only used the commit message but not the diff or only used the commit diff but not the message (\emph{commit information}),
let it generate the command line prompt in addition to the test input (\emph{task difficulty}),
used \gptmini/\deepseekr as less/more powerful LLMs or maximized the LLM temperature (\emph{LLM module}),
used ten (10) iterations instead of five (5) in the feedback loop (\emph{number of iterations}), or
dropping the execution analysis result from the feedback (\emph{feedback utility}).

\begin{table}
  \caption{\textbf{RQ2} results. Average effectiveness score on a slider representing effectiveness score $\in [0,3]$}
  \label{tab:rq2}
  \resizebox{\textwidth}{!}{%
    \begin{tabular}{@{}cc||c|ccc|ccc|cc||c|ccc|ccc|cc@{}}
      \toprule
      \multirow{4}{*}{Subject}              & \multirow{4}{*}{Issue}              & \multicolumn{9}{c||}{\textbf{Bug-Introducing-Commit (BIC)}} & \multicolumn{9}{c}{\textbf{Bug-Fixing-Commit (BFC)}}                                                                                                                                                                                                                                                                                                                                                                                                                                                                                                                                                                                                                                     \\ \cmidrule{3-20}
                                            &                                     & \multirow{3}{*}{\textsf{Default}}                    & \multicolumn{3}{c|}{Prompt Synthesizer}        & \multicolumn{3}{c|}{LLM Module}     & \multicolumn{2}{c||}{Exec. Analyzer} & \multirow{3}{*}{\textsf{Default}}   & \multicolumn{3}{c|}{Prompt Synthesizer} & \multicolumn{3}{c|}{LLM Module}     & \multicolumn{2}{c}{Exec. Analyzer}                                                                                                                                                                                                                                                                                                                                                                                              \\
                                            &                                     &                                                      & \textsf{\makecell{Only                                                                                                                                                                                                                                                                                                                                                                                                                                                                                                                                                                                                                                                             \\msg}}                       & \textsf{\makecell{Only\\diff}}              & \textsf{\makecell{Gen.\\cmd}}                      & \textsf{Temp\textsubscript{0.x}} & \textsf{\makecell{GPT4o\\mini}} & \textsf{\makecell{Deep-\\Seek R1}}            & \textsf{Iter\textsubscript{10}} & \textsf{\makecell{No\\feed.}} &                                   & \textsf{\makecell{Only\\msg}}                       & \textsf{\makecell{Only\\diff}}              & \textsf{\makecell{Gen.\\cmd}}                      & \textsf{Temp\textsubscript{0.x}} & \textsf{\makecell{GPT4o\\mini}} & \textsf{\makecell{Deep-\\Seek R1}}            & \textsf{Iter\textsubscript{10}} & \textsf{\makecell{No\\feed.}}            \\ \midrule
      \multirow{4}{*}{\mujs}                & \#65                                & \SLIDER{\mysliderlen}{1.00}{circle}                  & \SLIDER{\mysliderlen}{1.00}{circle}            & \SLIDER{\mysliderlen}{1.00}{circle} & \SLIDER{\mysliderlen}{1.00}{circle} & \SLIDER{\mysliderlen}{1.00}{circle} & \SLIDER{\mysliderlen}{0.20}{circle}     & \SLIDER{\mysliderlen}{1.00}{circle} & \SLIDER{\mysliderlen}{1.00}{circle} & \SLIDER{\mysliderlen}{0.90}{circle} & \SLIDER{\mysliderlen}{0.73}{circle} & \SLIDER{\mysliderlen}{0.33}{circle} & \SLIDER{\mysliderlen}{0.73}{circle} & \SLIDER{\mysliderlen}{0.73}{circle} & \SLIDER{\mysliderlen}{0.73}{circle} & \SLIDER{\mysliderlen}{0.33}{circle} & \SLIDER{\mysliderlen}{0.90}{circle} & \SLIDER{\mysliderlen}{0.73}{circle} & \SLIDER{\mysliderlen}{0.47}{circle} \\
                                            & \#141                               & \SLIDER{\mysliderlen}{0.67}{circle}                  & \SLIDER{\mysliderlen}{0.67}{circle}            & \SLIDER{\mysliderlen}{0.67}{circle} & \SLIDER{\mysliderlen}{0.67}{circle} & \SLIDER{\mysliderlen}{0.67}{circle} & \SLIDER{\mysliderlen}{0.67}{circle}     & \SLIDER{\mysliderlen}{0.67}{circle} & \SLIDER{\mysliderlen}{0.67}{circle} & \SLIDER{\mysliderlen}{0.67}{circle} & \SLIDER{\mysliderlen}{0.70}{circle} & \SLIDER{\mysliderlen}{0.00}{circle} & \SLIDER{\mysliderlen}{0.57}{circle} & \SLIDER{\mysliderlen}{0.70}{circle} & \SLIDER{\mysliderlen}{0.50}{circle} & \SLIDER{\mysliderlen}{0.70}{circle} & \SLIDER{\mysliderlen}{0.67}{circle} & \SLIDER{\mysliderlen}{0.63}{circle} & \SLIDER{\mysliderlen}{0.67}{circle} \\
                                            & \#145                               & \SLIDER{\mysliderlen}{0.87}{circle}                  & \SLIDER{\mysliderlen}{1.00}{circle}            & \SLIDER{\mysliderlen}{0.87}{circle} & \SLIDER{\mysliderlen}{0.87}{circle} & \SLIDER{\mysliderlen}{0.87}{circle} & \SLIDER{\mysliderlen}{0.40}{circle}     & \SLIDER{\mysliderlen}{1.00}{circle} & \SLIDER{\mysliderlen}{1.00}{circle} & \SLIDER{\mysliderlen}{0.73}{circle} & \SLIDER{\mysliderlen}{0.93}{circle} & \SLIDER{\mysliderlen}{0.80}{circle} & \SLIDER{\mysliderlen}{0.90}{circle} & \SLIDER{\mysliderlen}{0.93}{circle} & \SLIDER{\mysliderlen}{0.67}{circle} & \SLIDER{\mysliderlen}{0.27}{circle} & \SLIDER{\mysliderlen}{1.00}{circle} & \SLIDER{\mysliderlen}{1.00}{circle} & \SLIDER{\mysliderlen}{0.93}{circle} \\
                                            & \#166                               & \SLIDER{\mysliderlen}{0.37}{circle}                  & \SLIDER{\mysliderlen}{0.43}{circle}            & \SLIDER{\mysliderlen}{0.33}{circle} & \SLIDER{\mysliderlen}{0.37}{circle} & \SLIDER{\mysliderlen}{0.40}{circle} & \SLIDER{\mysliderlen}{0.33}{circle}     & \SLIDER{\mysliderlen}{0.60}{circle} & \SLIDER{\mysliderlen}{0.37}{circle} & \SLIDER{\mysliderlen}{0.37}{circle} & \SLIDER{\mysliderlen}{1.00}{circle} & \SLIDER{\mysliderlen}{1.00}{circle} & \SLIDER{\mysliderlen}{0.90}{circle} & \SLIDER{\mysliderlen}{1.00}{circle} & \SLIDER{\mysliderlen}{1.00}{circle} & \SLIDER{\mysliderlen}{1.00}{circle} & \SLIDER{\mysliderlen}{1.00}{circle} & \SLIDER{\mysliderlen}{1.00}{circle} & \SLIDER{\mysliderlen}{1.00}{circle} \\
      \midrule
      \multirow{5}{*}{\libxml} & \#535 & \SLIDER{\mysliderlen}{1.00}{circle} & \SLIDER{\mysliderlen}{1.00}{circle} & \SLIDER{\mysliderlen}{1.00}{circle} & \SLIDER{\mysliderlen}{0.40}{circle} & \SLIDER{\mysliderlen}{1.00}{circle} & \SLIDER{\mysliderlen}{1.00}{circle} & \SLIDER{\mysliderlen}{1.00}{circle} & \SLIDER{\mysliderlen}{1.00}{circle} & \SLIDER{\mysliderlen}{1.00}{circle} & \SLIDER{\mysliderlen}{1.00}{circle} & \SLIDER{\mysliderlen}{1.00}{circle} & \SLIDER{\mysliderlen}{1.00}{circle} & \SLIDER{\mysliderlen}{0.97}{circle} & \SLIDER{\mysliderlen}{1.00}{circle} & \SLIDER{\mysliderlen}{1.00}{circle} & \SLIDER{\mysliderlen}{1.00}{circle} & \SLIDER{\mysliderlen}{1.00}{circle} & \SLIDER{\mysliderlen}{1.00}{circle} \\
      & \#550 & \SLIDER{\mysliderlen}{0.40}{circle} & \SLIDER{\mysliderlen}{0.63}{circle} & \SLIDER{\mysliderlen}{0.37}{circle} & \SLIDER{\mysliderlen}{0.40}{circle} & \SLIDER{\mysliderlen}{0.50}{circle} & \SLIDER{\mysliderlen}{0.50}{circle} & \SLIDER{\mysliderlen}{0.67}{circle} & \SLIDER{\mysliderlen}{0.47}{circle} & \SLIDER{\mysliderlen}{0.37}{circle} & \SLIDER{\mysliderlen}{0.00}{circle} & \SLIDER{\mysliderlen}{0.00}{circle} & \SLIDER{\mysliderlen}{0.00}{circle} & \SLIDER{\mysliderlen}{0.00}{circle} & \SLIDER{\mysliderlen}{0.00}{circle} & \SLIDER{\mysliderlen}{0.00}{circle} & \SLIDER{\mysliderlen}{0.00}{circle} & \SLIDER{\mysliderlen}{0.00}{circle} & \SLIDER{\mysliderlen}{0.00}{circle} \\
      & \#553 & \SLIDER{\mysliderlen}{1.00}{circle} & \SLIDER{\mysliderlen}{1.00}{circle} & \SLIDER{\mysliderlen}{1.00}{circle} & \SLIDER{\mysliderlen}{0.33}{circle} & \SLIDER{\mysliderlen}{1.00}{circle} & \SLIDER{\mysliderlen}{1.00}{circle} & \SLIDER{\mysliderlen}{1.00}{circle} & \SLIDER{\mysliderlen}{1.00}{circle} & \SLIDER{\mysliderlen}{1.00}{circle} & \SLIDER{\mysliderlen}{1.00}{circle} & \SLIDER{\mysliderlen}{1.00}{circle} & \SLIDER{\mysliderlen}{1.00}{circle} & \SLIDER{\mysliderlen}{0.13}{circle} & \SLIDER{\mysliderlen}{1.00}{circle} & \SLIDER{\mysliderlen}{1.00}{circle} & \SLIDER{\mysliderlen}{1.00}{circle} & \SLIDER{\mysliderlen}{1.00}{circle} & \SLIDER{\mysliderlen}{1.00}{circle} \\
      & \#720 & \SLIDER{\mysliderlen}{0.33}{circle} & \SLIDER{\mysliderlen}{0.33}{circle} & \SLIDER{\mysliderlen}{0.33}{circle} & \SLIDER{\mysliderlen}{0.37}{circle} & \SLIDER{\mysliderlen}{0.33}{circle} & \SLIDER{\mysliderlen}{0.33}{circle} & \SLIDER{\mysliderlen}{0.37}{circle} & \SLIDER{\mysliderlen}{0.33}{circle} & \SLIDER{\mysliderlen}{0.33}{circle} & \SLIDER{\mysliderlen}{0.23}{circle} & \SLIDER{\mysliderlen}{0.33}{circle} & \SLIDER{\mysliderlen}{0.10}{circle} & \SLIDER{\mysliderlen}{0.03}{circle} & \SLIDER{\mysliderlen}{0.10}{circle} & \SLIDER{\mysliderlen}{0.27}{circle} & \SLIDER{\mysliderlen}{0.33}{circle} & \SLIDER{\mysliderlen}{0.30}{circle} & \SLIDER{\mysliderlen}{0.17}{circle} \\
      & \#841 & \SLIDER{\mysliderlen}{1.00}{circle} & \SLIDER{\mysliderlen}{1.00}{circle} & \SLIDER{\mysliderlen}{1.00}{circle} & \SLIDER{\mysliderlen}{0.67}{circle} & \SLIDER{\mysliderlen}{1.00}{circle} & \SLIDER{\mysliderlen}{1.00}{circle} & \SLIDER{\mysliderlen}{1.00}{circle} & \SLIDER{\mysliderlen}{1.00}{circle} & \SLIDER{\mysliderlen}{1.00}{circle} & \SLIDER{\mysliderlen}{1.00}{circle} & \SLIDER{\mysliderlen}{1.00}{circle} & \SLIDER{\mysliderlen}{1.00}{circle} & \SLIDER{\mysliderlen}{1.00}{circle} & \SLIDER{\mysliderlen}{1.00}{circle} & \SLIDER{\mysliderlen}{1.00}{circle} & \SLIDER{\mysliderlen}{1.00}{circle} & \SLIDER{\mysliderlen}{1.00}{circle} & \SLIDER{\mysliderlen}{1.00}{circle} \\
      \midrule
      \multirow{5}{*}{\poppler}             & \#1282                              & \SLIDER{\mysliderlen}{0.37}{circle}                  & \SLIDER{\mysliderlen}{0.43}{circle}            & \SLIDER{\mysliderlen}{0.33}{circle} & \SLIDER{\mysliderlen}{0.33}{circle} & \SLIDER{\mysliderlen}{0.47}{circle} & \SLIDER{\mysliderlen}{0.03}{circle}     & \SLIDER{\mysliderlen}{0.53}{circle} & \SLIDER{\mysliderlen}{0.40}{circle} & \SLIDER{\mysliderlen}{0.37}{circle} & \SLIDER{\mysliderlen}{0.30}{circle} & \SLIDER{\mysliderlen}{0.33}{circle} & \SLIDER{\mysliderlen}{0.30}{circle} & \SLIDER{\mysliderlen}{0.33}{circle} & \SLIDER{\mysliderlen}{0.33}{circle} & \SLIDER{\mysliderlen}{0.10}{circle} & \SLIDER{\mysliderlen}{0.07}{circle} & \SLIDER{\mysliderlen}{0.33}{circle} & \SLIDER{\mysliderlen}{0.30}{circle} \\
                                            & \#1289                              & \SLIDER{\mysliderlen}{0.13}{circle}                  & \SLIDER{\mysliderlen}{0.20}{circle}            & \SLIDER{\mysliderlen}{0.33}{circle} & \SLIDER{\mysliderlen}{0.30}{circle} & \SLIDER{\mysliderlen}{0.50}{circle} & \SLIDER{\mysliderlen}{0.13}{circle}     & \SLIDER{\mysliderlen}{0.60}{circle} & \SLIDER{\mysliderlen}{0.20}{circle} & \SLIDER{\mysliderlen}{0.00}{circle} & \SLIDER{\mysliderlen}{0.67}{circle} & \SLIDER{\mysliderlen}{0.00}{circle} & \SLIDER{\mysliderlen}{0.50}{circle} & \SLIDER{\mysliderlen}{0.00}{circle} & \SLIDER{\mysliderlen}{0.70}{circle} & \SLIDER{\mysliderlen}{0.00}{circle} & \SLIDER{\mysliderlen}{1.00}{circle} & \SLIDER{\mysliderlen}{0.87}{circle} & \SLIDER{\mysliderlen}{0.67}{circle} \\
                                            & \#1303                              & \SLIDER{\mysliderlen}{0.00}{circle}                  & \SLIDER{\mysliderlen}{0.00}{circle}            & \SLIDER{\mysliderlen}{0.00}{circle} & \SLIDER{\mysliderlen}{0.00}{circle} & \SLIDER{\mysliderlen}{0.00}{circle} & \SLIDER{\mysliderlen}{0.00}{circle}     & \SLIDER{\mysliderlen}{0.00}{circle} & \SLIDER{\mysliderlen}{0.00}{circle} & \SLIDER{\mysliderlen}{0.00}{circle} & \SLIDER{\mysliderlen}{0.07}{circle} & \SLIDER{\mysliderlen}{0.00}{circle} & \SLIDER{\mysliderlen}{0.23}{circle} & \SLIDER{\mysliderlen}{0.00}{circle} & \SLIDER{\mysliderlen}{0.40}{circle} & \SLIDER{\mysliderlen}{0.00}{circle} & \SLIDER{\mysliderlen}{0.60}{circle} & \SLIDER{\mysliderlen}{0.33}{circle} & \SLIDER{\mysliderlen}{0.00}{circle} \\
                                            & \#1305                              & \SLIDER{\mysliderlen}{0.00}{circle}                  & \SLIDER{\mysliderlen}{0.00}{circle}            & \SLIDER{\mysliderlen}{0.00}{circle} & \SLIDER{\mysliderlen}{0.00}{circle} & \SLIDER{\mysliderlen}{0.00}{circle} & \SLIDER{\mysliderlen}{0.00}{circle}     & \SLIDER{\mysliderlen}{0.40}{circle} & \SLIDER{\mysliderlen}{0.00}{circle} & \SLIDER{\mysliderlen}{0.00}{circle} & \SLIDER{\mysliderlen}{0.00}{circle} & \SLIDER{\mysliderlen}{0.00}{circle} & \SLIDER{\mysliderlen}{0.00}{circle} & \SLIDER{\mysliderlen}{0.00}{circle} & \SLIDER{\mysliderlen}{0.00}{circle} & \SLIDER{\mysliderlen}{0.00}{circle} & \SLIDER{\mysliderlen}{0.00}{circle} & \SLIDER{\mysliderlen}{0.00}{circle} & \SLIDER{\mysliderlen}{0.00}{circle} \\
                                            & \#1381                              & \SLIDER{\mysliderlen}{0.00}{circle}                  & \SLIDER{\mysliderlen}{0.00}{circle}            & \SLIDER{\mysliderlen}{0.00}{circle} & \SLIDER{\mysliderlen}{0.00}{circle} & \SLIDER{\mysliderlen}{0.00}{circle} & \SLIDER{\mysliderlen}{0.00}{circle}     & \SLIDER{\mysliderlen}{0.00}{circle} & \SLIDER{\mysliderlen}{0.00}{circle} & \SLIDER{\mysliderlen}{0.00}{circle} & \SLIDER{\mysliderlen}{0.00}{circle} & \SLIDER{\mysliderlen}{0.00}{circle} & \SLIDER{\mysliderlen}{0.00}{circle} & \SLIDER{\mysliderlen}{0.00}{circle} & \SLIDER{\mysliderlen}{0.00}{circle} & \SLIDER{\mysliderlen}{0.00}{circle} & \SLIDER{\mysliderlen}{0.03}{circle} & \SLIDER{\mysliderlen}{0.00}{circle} & \SLIDER{\mysliderlen}{0.00}{circle} \\
      \midrule
      \multirow{4}{*}{\jerryscript}         & \#5013                              & \SLIDER{\mysliderlen}{0.33}{circle}                  & \SLIDER{\mysliderlen}{0.33}{circle}            & \SLIDER{\mysliderlen}{0.33}{circle} & \SLIDER{\mysliderlen}{0.33}{circle} & \SLIDER{\mysliderlen}{0.37}{circle} & \SLIDER{\mysliderlen}{0.03}{circle}     & \SLIDER{\mysliderlen}{0.57}{circle} & \SLIDER{\mysliderlen}{0.33}{circle} & \SLIDER{\mysliderlen}{0.33}{circle} & \SLIDER{\mysliderlen}{0.33}{circle} & \SLIDER{\mysliderlen}{0.33}{circle} & \SLIDER{\mysliderlen}{0.00}{circle} & \SLIDER{\mysliderlen}{0.33}{circle} & \SLIDER{\mysliderlen}{0.33}{circle} & \SLIDER{\mysliderlen}{0.33}{circle} & \SLIDER{\mysliderlen}{0.93}{circle} & \SLIDER{\mysliderlen}{0.33}{circle} & \SLIDER{\mysliderlen}{0.30}{circle} \\
                                            & \#5117                              & \SLIDER{\mysliderlen}{0.67}{circle}                  & \SLIDER{\mysliderlen}{0.67}{circle}            & \SLIDER{\mysliderlen}{0.67}{circle} & \SLIDER{\mysliderlen}{0.67}{circle} & \SLIDER{\mysliderlen}{0.67}{circle} & \SLIDER{\mysliderlen}{0.67}{circle}     & \SLIDER{\mysliderlen}{0.67}{circle} & \SLIDER{\mysliderlen}{0.67}{circle} & \SLIDER{\mysliderlen}{0.67}{circle} & \SLIDER{\mysliderlen}{0.67}{circle} & \SLIDER{\mysliderlen}{0.47}{circle} & \SLIDER{\mysliderlen}{0.33}{circle} & \SLIDER{\mysliderlen}{0.67}{circle} & \SLIDER{\mysliderlen}{0.73}{circle} & \SLIDER{\mysliderlen}{0.33}{circle} & \SLIDER{\mysliderlen}{0.53}{circle} & \SLIDER{\mysliderlen}{0.53}{circle} & \SLIDER{\mysliderlen}{0.67}{circle} \\
                                            & \#5138                              & \SLIDER{\mysliderlen}{0.73}{circle}                  & \SLIDER{\mysliderlen}{0.67}{circle}            & \SLIDER{\mysliderlen}{0.77}{circle} & \SLIDER{\mysliderlen}{0.73}{circle} & \SLIDER{\mysliderlen}{0.73}{circle} & \SLIDER{\mysliderlen}{0.67}{circle}     & \SLIDER{\mysliderlen}{0.67}{circle} & \SLIDER{\mysliderlen}{0.80}{circle} & \SLIDER{\mysliderlen}{0.77}{circle} & \SLIDER{\mysliderlen}{0.33}{circle} & \SLIDER{\mysliderlen}{0.33}{circle} & \SLIDER{\mysliderlen}{0.33}{circle} & \SLIDER{\mysliderlen}{0.33}{circle} & \SLIDER{\mysliderlen}{0.53}{circle} & \SLIDER{\mysliderlen}{0.17}{circle} & \SLIDER{\mysliderlen}{1.00}{circle} & \SLIDER{\mysliderlen}{0.47}{circle} & \SLIDER{\mysliderlen}{0.50}{circle} \\
                                            & \#5153                              & \SLIDER{\mysliderlen}{0.67}{circle}                  & \SLIDER{\mysliderlen}{0.67}{circle}            & \SLIDER{\mysliderlen}{0.77}{circle} & \SLIDER{\mysliderlen}{0.67}{circle} & \SLIDER{\mysliderlen}{0.70}{circle} & \SLIDER{\mysliderlen}{0.67}{circle}     & \SLIDER{\mysliderlen}{0.67}{circle} & \SLIDER{\mysliderlen}{0.77}{circle} & \SLIDER{\mysliderlen}{0.73}{circle} & \SLIDER{\mysliderlen}{0.00}{circle} & \SLIDER{\mysliderlen}{0.00}{circle} & \SLIDER{\mysliderlen}{0.00}{circle} & \SLIDER{\mysliderlen}{0.00}{circle} & \SLIDER{\mysliderlen}{0.00}{circle} & \SLIDER{\mysliderlen}{0.00}{circle} & \SLIDER{\mysliderlen}{0.00}{circle} & \SLIDER{\mysliderlen}{0.00}{circle} & \SLIDER{\mysliderlen}{0.00}{circle} \\
      \midrule
      \multirow{4}{*}{\zthree}              & \#6659                              & \SLIDER{\mysliderlen}{0.00}{circle}                  & \SLIDER{\mysliderlen}{0.00}{circle}            & \SLIDER{\mysliderlen}{0.00}{circle} & \SLIDER{\mysliderlen}{0.00}{circle} & \SLIDER{\mysliderlen}{0.00}{circle} & \SLIDER{\mysliderlen}{0.00}{circle}     & \SLIDER{\mysliderlen}{0.00}{circle} & \SLIDER{\mysliderlen}{0.00}{circle} & \SLIDER{\mysliderlen}{0.00}{circle} & \SLIDER{\mysliderlen}{0.00}{circle} & \SLIDER{\mysliderlen}{0.00}{circle} & \SLIDER{\mysliderlen}{0.00}{circle} & \SLIDER{\mysliderlen}{0.00}{circle} & \SLIDER{\mysliderlen}{0.07}{circle} & \SLIDER{\mysliderlen}{0.00}{circle} & \SLIDER{\mysliderlen}{0.00}{circle} & \SLIDER{\mysliderlen}{0.00}{circle} & \SLIDER{\mysliderlen}{0.00}{circle} \\
                                            & \#6914                              & \SLIDER{\mysliderlen}{0.07}{circle}                  & \SLIDER{\mysliderlen}{0.03}{circle}            & \SLIDER{\mysliderlen}{0.13}{circle} & \SLIDER{\mysliderlen}{0.07}{circle} & \SLIDER{\mysliderlen}{0.13}{circle} & \SLIDER{\mysliderlen}{0.07}{circle}     & \SLIDER{\mysliderlen}{0.10}{circle} & \SLIDER{\mysliderlen}{0.03}{circle} & \SLIDER{\mysliderlen}{0.03}{circle} & \SLIDER{\mysliderlen}{0.00}{circle} & \SLIDER{\mysliderlen}{0.07}{circle} & \SLIDER{\mysliderlen}{0.00}{circle} & \SLIDER{\mysliderlen}{0.00}{circle} & \SLIDER{\mysliderlen}{0.10}{circle} & \SLIDER{\mysliderlen}{0.00}{circle} & \SLIDER{\mysliderlen}{0.17}{circle} & \SLIDER{\mysliderlen}{0.03}{circle} & \SLIDER{\mysliderlen}{0.00}{circle} \\
                                            & \#7246                              & \SLIDER{\mysliderlen}{0.33}{circle}                  & \SLIDER{\mysliderlen}{0.33}{circle}            & \SLIDER{\mysliderlen}{0.33}{circle} & \SLIDER{\mysliderlen}{0.33}{circle} & \SLIDER{\mysliderlen}{0.33}{circle} & \SLIDER{\mysliderlen}{0.33}{circle}     & \SLIDER{\mysliderlen}{0.67}{circle} & \SLIDER{\mysliderlen}{0.33}{circle} & \SLIDER{\mysliderlen}{0.33}{circle} & \SLIDER{\mysliderlen}{0.00}{circle} & \SLIDER{\mysliderlen}{0.00}{circle} & \SLIDER{\mysliderlen}{0.07}{circle} & \SLIDER{\mysliderlen}{0.00}{circle} & \SLIDER{\mysliderlen}{0.00}{circle} & \SLIDER{\mysliderlen}{0.13}{circle} & \SLIDER{\mysliderlen}{0.00}{circle} & \SLIDER{\mysliderlen}{0.00}{circle} & \SLIDER{\mysliderlen}{0.00}{circle} \\
                                            & \#7252                              & \SLIDER{\mysliderlen}{0.27}{circle}                  & \SLIDER{\mysliderlen}{0.00}{circle}            & \SLIDER{\mysliderlen}{0.30}{circle} & \SLIDER{\mysliderlen}{0.27}{circle} & \SLIDER{\mysliderlen}{0.13}{circle} & \SLIDER{\mysliderlen}{0.00}{circle}     & \SLIDER{\mysliderlen}{0.33}{circle} & \SLIDER{\mysliderlen}{0.23}{circle} & \SLIDER{\mysliderlen}{0.03}{circle} & \SLIDER{\mysliderlen}{0.17}{circle} & \SLIDER{\mysliderlen}{0.00}{circle} & \SLIDER{\mysliderlen}{0.20}{circle} & \SLIDER{\mysliderlen}{0.17}{circle} & \SLIDER{\mysliderlen}{0.13}{circle} & \SLIDER{\mysliderlen}{0.00}{circle} & \SLIDER{\mysliderlen}{0.33}{circle} & \SLIDER{\mysliderlen}{0.27}{circle} & \SLIDER{\mysliderlen}{0.07}{circle} \\
      \midrule
      \multirow{4}{*}{\php}                 & \#16777                             & \SLIDER{\mysliderlen}{0.33}{circle}                  & \SLIDER{\mysliderlen}{0.20}{circle}            & \SLIDER{\mysliderlen}{0.30}{circle} & \SLIDER{\mysliderlen}{0.33}{circle} & \SLIDER{\mysliderlen}{0.33}{circle} & \SLIDER{\mysliderlen}{0.30}{circle}     & \SLIDER{\mysliderlen}{0.33}{circle} & \SLIDER{\mysliderlen}{0.33}{circle} & \SLIDER{\mysliderlen}{0.33}{circle} & \SLIDER{\mysliderlen}{0.33}{circle} & \SLIDER{\mysliderlen}{0.20}{circle} & \SLIDER{\mysliderlen}{0.27}{circle} & \SLIDER{\mysliderlen}{0.33}{circle} & \SLIDER{\mysliderlen}{0.30}{circle} & \SLIDER{\mysliderlen}{0.17}{circle} & \SLIDER{\mysliderlen}{0.33}{circle} & \SLIDER{\mysliderlen}{0.30}{circle} & \SLIDER{\mysliderlen}{0.10}{circle} \\
                                            & \#16978                             & \SLIDER{\mysliderlen}{0.37}{circle}                  & \SLIDER{\mysliderlen}{0.50}{circle}            & \SLIDER{\mysliderlen}{0.43}{circle} & \SLIDER{\mysliderlen}{0.37}{circle} & \SLIDER{\mysliderlen}{0.43}{circle} & \SLIDER{\mysliderlen}{0.33}{circle}     & \SLIDER{\mysliderlen}{0.40}{circle} & \SLIDER{\mysliderlen}{0.43}{circle} & \SLIDER{\mysliderlen}{0.40}{circle} & \SLIDER{\mysliderlen}{0.00}{circle} & \SLIDER{\mysliderlen}{0.00}{circle} & \SLIDER{\mysliderlen}{0.03}{circle} & \SLIDER{\mysliderlen}{0.00}{circle} & \SLIDER{\mysliderlen}{0.00}{circle} & \SLIDER{\mysliderlen}{0.00}{circle} & \SLIDER{\mysliderlen}{0.03}{circle} & \SLIDER{\mysliderlen}{0.00}{circle} & \SLIDER{\mysliderlen}{0.00}{circle} \\
                                            & \#17442                             & \SLIDER{\mysliderlen}{0.67}{circle}                  & \SLIDER{\mysliderlen}{0.57}{circle}            & \SLIDER{\mysliderlen}{0.67}{circle} & \SLIDER{\mysliderlen}{0.67}{circle} & \SLIDER{\mysliderlen}{0.60}{circle} & \SLIDER{\mysliderlen}{0.60}{circle}     & \SLIDER{\mysliderlen}{0.67}{circle} & \SLIDER{\mysliderlen}{0.67}{circle} & \SLIDER{\mysliderlen}{0.67}{circle} & \SLIDER{\mysliderlen}{0.00}{circle} & \SLIDER{\mysliderlen}{0.00}{circle} & \SLIDER{\mysliderlen}{0.00}{circle} & \SLIDER{\mysliderlen}{0.00}{circle} & \SLIDER{\mysliderlen}{0.00}{circle} & \SLIDER{\mysliderlen}{0.00}{circle} & \SLIDER{\mysliderlen}{0.00}{circle} & \SLIDER{\mysliderlen}{0.00}{circle} & \SLIDER{\mysliderlen}{0.00}{circle} \\
                                            & \#17463                             & \SLIDER{\mysliderlen}{0.20}{circle}                  & \SLIDER{\mysliderlen}{0.33}{circle}            & \SLIDER{\mysliderlen}{0.10}{circle} & \SLIDER{\mysliderlen}{0.20}{circle} & \SLIDER{\mysliderlen}{0.23}{circle} & \SLIDER{\mysliderlen}{0.10}{circle}     & \SLIDER{\mysliderlen}{0.30}{circle} & \SLIDER{\mysliderlen}{0.27}{circle} & \SLIDER{\mysliderlen}{0.00}{circle} & \SLIDER{\mysliderlen}{1.00}{circle} & \SLIDER{\mysliderlen}{1.00}{circle} & \SLIDER{\mysliderlen}{0.67}{circle} & \SLIDER{\mysliderlen}{1.00}{circle} & \SLIDER{\mysliderlen}{1.00}{circle} & \SLIDER{\mysliderlen}{1.00}{circle} & \SLIDER{\mysliderlen}{1.00}{circle} & \SLIDER{\mysliderlen}{1.00}{circle} & \SLIDER{\mysliderlen}{1.00}{circle} \\
      \midrule
      \multirow{4}{*}{\jq} & \#2825 & \SLIDER{\mysliderlen}{0.33}{circle} & \SLIDER{\mysliderlen}{0.40}{circle} & \SLIDER{\mysliderlen}{0.33}{circle} & \SLIDER{\mysliderlen}{0.33}{circle} & \SLIDER{\mysliderlen}{0.33}{circle} & \SLIDER{\mysliderlen}{0.33}{circle} & \SLIDER{\mysliderlen}{0.37}{circle} & \SLIDER{\mysliderlen}{0.33}{circle} & \SLIDER{\mysliderlen}{0.33}{circle} & \SLIDER{\mysliderlen}{0.37}{circle} & \SLIDER{\mysliderlen}{0.33}{circle} & \SLIDER{\mysliderlen}{0.43}{circle} & \SLIDER{\mysliderlen}{0.47}{circle} & \SLIDER{\mysliderlen}{0.40}{circle} & \SLIDER{\mysliderlen}{0.33}{circle} & \SLIDER{\mysliderlen}{0.40}{circle} & \SLIDER{\mysliderlen}{0.37}{circle} & \SLIDER{\mysliderlen}{0.33}{circle} \\
      & \#2976 & \SLIDER{\mysliderlen}{0.33}{circle} & \SLIDER{\mysliderlen}{0.33}{circle} & \SLIDER{\mysliderlen}{0.33}{circle} & \SLIDER{\mysliderlen}{0.37}{circle} & \SLIDER{\mysliderlen}{0.33}{circle} & \SLIDER{\mysliderlen}{0.33}{circle} & \SLIDER{\mysliderlen}{0.33}{circle} & \SLIDER{\mysliderlen}{0.33}{circle} & \SLIDER{\mysliderlen}{0.33}{circle} & \SLIDER{\mysliderlen}{1.00}{circle} & \SLIDER{\mysliderlen}{1.00}{circle} & \SLIDER{\mysliderlen}{0.33}{circle} & \SLIDER{\mysliderlen}{1.00}{circle} & \SLIDER{\mysliderlen}{1.00}{circle} & \SLIDER{\mysliderlen}{1.00}{circle} & \SLIDER{\mysliderlen}{1.00}{circle} & \SLIDER{\mysliderlen}{1.00}{circle} & \SLIDER{\mysliderlen}{1.00}{circle} \\
      & \#3262 & \SLIDER{\mysliderlen}{1.00}{circle} & \SLIDER{\mysliderlen}{1.00}{circle} & \SLIDER{\mysliderlen}{1.00}{circle} & \SLIDER{\mysliderlen}{0.37}{circle} & \SLIDER{\mysliderlen}{1.00}{circle} & \SLIDER{\mysliderlen}{0.00}{circle} & \SLIDER{\mysliderlen}{1.00}{circle} & \SLIDER{\mysliderlen}{1.00}{circle} & \SLIDER{\mysliderlen}{0.10}{circle} & \SLIDER{\mysliderlen}{1.00}{circle} & \SLIDER{\mysliderlen}{0.93}{circle} & \SLIDER{\mysliderlen}{1.00}{circle} & \SLIDER{\mysliderlen}{0.37}{circle} & \SLIDER{\mysliderlen}{1.00}{circle} & \SLIDER{\mysliderlen}{0.67}{circle} & \SLIDER{\mysliderlen}{1.00}{circle} & \SLIDER{\mysliderlen}{1.00}{circle} & \SLIDER{\mysliderlen}{1.00}{circle} \\
      & \#49014 & \SLIDER{\mysliderlen}{0.13}{circle} & \SLIDER{\mysliderlen}{0.00}{circle} & \SLIDER{\mysliderlen}{0.13}{circle} & \SLIDER{\mysliderlen}{0.07}{circle} & \SLIDER{\mysliderlen}{0.20}{circle} & \SLIDER{\mysliderlen}{0.00}{circle} & \SLIDER{\mysliderlen}{0.43}{circle} & \SLIDER{\mysliderlen}{0.17}{circle} & \SLIDER{\mysliderlen}{0.07}{circle} & \SLIDER{\mysliderlen}{0.00}{circle} & \SLIDER{\mysliderlen}{0.00}{circle} & \SLIDER{\mysliderlen}{0.00}{circle} & \SLIDER{\mysliderlen}{0.00}{circle} & \SLIDER{\mysliderlen}{0.03}{circle} & \SLIDER{\mysliderlen}{0.00}{circle} & \SLIDER{\mysliderlen}{0.10}{circle} & \SLIDER{\mysliderlen}{0.00}{circle} & \SLIDER{\mysliderlen}{0.00}{circle} \\
      \midrule
      \multirow{6}{*}{\micropython} & \#13007 & \SLIDER{\mysliderlen}{0.53}{circle} & \SLIDER{\mysliderlen}{0.77}{circle} & \SLIDER{\mysliderlen}{0.37}{circle} & \SLIDER{\mysliderlen}{0.53}{circle} & \SLIDER{\mysliderlen}{0.50}{circle} & \SLIDER{\mysliderlen}{0.00}{circle} & \SLIDER{\mysliderlen}{0.50}{circle} & \SLIDER{\mysliderlen}{0.63}{circle} & \SLIDER{\mysliderlen}{0.00}{circle} & \SLIDER{\mysliderlen}{0.07}{circle} & \SLIDER{\mysliderlen}{0.17}{circle} & \SLIDER{\mysliderlen}{0.23}{circle} & \SLIDER{\mysliderlen}{0.07}{circle} & \SLIDER{\mysliderlen}{0.00}{circle} & \SLIDER{\mysliderlen}{0.00}{circle} & \SLIDER{\mysliderlen}{1.00}{circle} & \SLIDER{\mysliderlen}{0.10}{circle} & \SLIDER{\mysliderlen}{0.00}{circle} \\
      & \#13041 & \SLIDER{\mysliderlen}{0.00}{circle} & \SLIDER{\mysliderlen}{0.10}{circle} & \SLIDER{\mysliderlen}{0.10}{circle} & \SLIDER{\mysliderlen}{0.00}{circle} & \SLIDER{\mysliderlen}{0.00}{circle} & \SLIDER{\mysliderlen}{0.00}{circle} & \SLIDER{\mysliderlen}{0.00}{circle} & \SLIDER{\mysliderlen}{0.00}{circle} & \SLIDER{\mysliderlen}{0.00}{circle} & \SLIDER{\mysliderlen}{0.67}{circle} & \SLIDER{\mysliderlen}{0.70}{circle} & \SLIDER{\mysliderlen}{0.67}{circle} & \SLIDER{\mysliderlen}{0.67}{circle} & \SLIDER{\mysliderlen}{0.67}{circle} & \SLIDER{\mysliderlen}{0.67}{circle} & \SLIDER{\mysliderlen}{0.70}{circle} & \SLIDER{\mysliderlen}{0.67}{circle} & \SLIDER{\mysliderlen}{0.70}{circle} \\
      & \#17733 & \SLIDER{\mysliderlen}{0.33}{circle} & \SLIDER{\mysliderlen}{0.33}{circle} & \SLIDER{\mysliderlen}{0.33}{circle} & \SLIDER{\mysliderlen}{0.33}{circle} & \SLIDER{\mysliderlen}{0.33}{circle} & \SLIDER{\mysliderlen}{0.33}{circle} & \SLIDER{\mysliderlen}{0.33}{circle} & \SLIDER{\mysliderlen}{0.37}{circle} & \SLIDER{\mysliderlen}{0.33}{circle} & \SLIDER{\mysliderlen}{0.33}{circle} & \SLIDER{\mysliderlen}{0.33}{circle} & \SLIDER{\mysliderlen}{0.33}{circle} & \SLIDER{\mysliderlen}{0.33}{circle} & \SLIDER{\mysliderlen}{0.33}{circle} & \SLIDER{\mysliderlen}{0.33}{circle} & \SLIDER{\mysliderlen}{0.33}{circle} & \SLIDER{\mysliderlen}{0.33}{circle} & \SLIDER{\mysliderlen}{0.30}{circle} \\
      & \#17815 & \SLIDER{\mysliderlen}{0.63}{circle} & \SLIDER{\mysliderlen}{0.60}{circle} & \SLIDER{\mysliderlen}{0.37}{circle} & \SLIDER{\mysliderlen}{0.63}{circle} & \SLIDER{\mysliderlen}{0.67}{circle} & \SLIDER{\mysliderlen}{0.63}{circle} & \SLIDER{\mysliderlen}{0.60}{circle} & \SLIDER{\mysliderlen}{0.67}{circle} & \SLIDER{\mysliderlen}{0.67}{circle} & \SLIDER{\mysliderlen}{1.00}{circle} & \SLIDER{\mysliderlen}{1.00}{circle} & \SLIDER{\mysliderlen}{0.13}{circle} & \SLIDER{\mysliderlen}{1.00}{circle} & \SLIDER{\mysliderlen}{1.00}{circle} & \SLIDER{\mysliderlen}{1.00}{circle} & \SLIDER{\mysliderlen}{1.00}{circle} & \SLIDER{\mysliderlen}{1.00}{circle} & \SLIDER{\mysliderlen}{1.00}{circle} \\
      & \#17841 & \SLIDER{\mysliderlen}{0.67}{circle} & \SLIDER{\mysliderlen}{0.67}{circle} & \SLIDER{\mysliderlen}{0.67}{circle} & \SLIDER{\mysliderlen}{0.67}{circle} & \SLIDER{\mysliderlen}{0.77}{circle} & \SLIDER{\mysliderlen}{0.63}{circle} & \SLIDER{\mysliderlen}{0.67}{circle} & \SLIDER{\mysliderlen}{0.70}{circle} & \SLIDER{\mysliderlen}{0.73}{circle} & \SLIDER{\mysliderlen}{0.33}{circle} & \SLIDER{\mysliderlen}{0.33}{circle} & \SLIDER{\mysliderlen}{0.33}{circle} & \SLIDER{\mysliderlen}{0.33}{circle} & \SLIDER{\mysliderlen}{0.33}{circle} & \SLIDER{\mysliderlen}{0.33}{circle} & \SLIDER{\mysliderlen}{0.40}{circle} & \SLIDER{\mysliderlen}{0.33}{circle} & \SLIDER{\mysliderlen}{0.33}{circle} \\
      & \#17847 & \SLIDER{\mysliderlen}{0.67}{circle} & \SLIDER{\mysliderlen}{0.67}{circle} & \SLIDER{\mysliderlen}{0.67}{circle} & \SLIDER{\mysliderlen}{0.67}{circle} & \SLIDER{\mysliderlen}{0.67}{circle} & \SLIDER{\mysliderlen}{0.00}{circle} & \SLIDER{\mysliderlen}{1.00}{circle} & \SLIDER{\mysliderlen}{0.67}{circle} & \SLIDER{\mysliderlen}{0.67}{circle} & \SLIDER{\mysliderlen}{0.00}{circle} & \SLIDER{\mysliderlen}{0.00}{circle} & \SLIDER{\mysliderlen}{0.00}{circle} & \SLIDER{\mysliderlen}{0.00}{circle} & \SLIDER{\mysliderlen}{0.00}{circle} & \SLIDER{\mysliderlen}{0.00}{circle} & \SLIDER{\mysliderlen}{0.63}{circle} & \SLIDER{\mysliderlen}{0.00}{circle} & \SLIDER{\mysliderlen}{0.00}{circle} \\
      \midrule
      \multicolumn{2}{c||}{\textbf{Average}} &
      \begin{tabular}{c}\SLIDER{\mysliderlen}{0.46}{circle}\vspace{-0.1cm}\\\small{1.37}\end{tabular} & 
      \begin{tabular}{c}\SLIDER{\mysliderlen}{0.47}{circle}\vspace{-0.1cm}\\\small{1.41}\end{tabular} & 
      \begin{tabular}{c}\SLIDER{\mysliderlen}{0.45}{circle}\vspace{-0.1cm}\\\small{1.36}\end{tabular} & 
      \begin{tabular}{c}\SLIDER{\mysliderlen}{0.40}{circle}\vspace{-0.1cm}\\\small{1.19}\end{tabular} & 
      \begin{tabular}{c}\SLIDER{\mysliderlen}{0.48}{circle}\vspace{-0.1cm}\\\small{1.43}\end{tabular} & 
      \begin{tabular}{c}\SLIDER{\mysliderlen}{0.32}{circle}\vspace{-0.1cm}\\\small{0.97}\end{tabular} & 
      \begin{tabular}{c}\SLIDER{\mysliderlen}{0.54}{circle}\vspace{-0.1cm}\\\small{1.62}\end{tabular} & 
      \begin{tabular}{c}\SLIDER{\mysliderlen}{0.48}{circle}\vspace{-0.1cm}\\\small{1.43}\end{tabular} & 
      \begin{tabular}{c}\SLIDER{\mysliderlen}{0.40}{circle}\vspace{-0.1cm}\\\small{1.19}\end{tabular} & 
      \begin{tabular}{c}\SLIDER{\mysliderlen}{0.42}{circle}\vspace{-0.1cm}\\\small{1.27}\end{tabular} & 
      \begin{tabular}{c}\SLIDER{\mysliderlen}{0.36}{circle}\vspace{-0.1cm}\\\small{1.08}\end{tabular} & 
      \begin{tabular}{c}\SLIDER{\mysliderlen}{0.35}{circle}\vspace{-0.1cm}\\\small{1.05}\end{tabular} & 
      \begin{tabular}{c}\SLIDER{\mysliderlen}{0.36}{circle}\vspace{-0.1cm}\\\small{1.08}\end{tabular} & 
      \begin{tabular}{c}\SLIDER{\mysliderlen}{0.43}{circle}\vspace{-0.1cm}\\\small{1.28}\end{tabular} & 
      \begin{tabular}{c}\SLIDER{\mysliderlen}{0.34}{circle}\vspace{-0.1cm}\\\small{1.01}\end{tabular} & 
      \begin{tabular}{c}\SLIDER{\mysliderlen}{0.54}{circle}\vspace{-0.1cm}\\\small{1.62}\end{tabular} & 
      \begin{tabular}{c}\SLIDER{\mysliderlen}{0.44}{circle}\vspace{-0.1cm}\\\small{1.33}\end{tabular} & 
      \begin{tabular}{c}\SLIDER{\mysliderlen}{0.40}{circle}\vspace{-0.1cm}\\\small{1.21}\end{tabular} \\
      \bottomrule
    \end{tabular}}
\end{table}

\textbf{Commit Information}.
We evaluate the impact of only using the commit message but not the diff (\textsf{Only msg}) or only using the commit diff but not the message (\textsf{Only diff}) and find that the results reproduce as long as the \emph{intention} of the change is captured.
\review{  %
In the bug-finding scenario, providing only the commit message or diff only changes the score slightly from \review{1.37} (meaning `sometimes output-changing') to \review{1.41} and \review{1.36}, respectively.
However, in the bug-reproduction scenario, we observe cases where a single source is insufficient: some commit messages lack detail while others are too high-level compared to the diff.
Specifically, some messages are too brief (e.g., ``\textsf{Fix \#IssueId}'') and provide less information than the code changes (e.g., Issue~\#141 in \mujs, where the diff details specific escape sequences absent from its message).
In contrast, other messages provide critical high-level intent (e.g.,  \micropython Issue~\#13087 ``\textsf{Fix int.to\_bytes() buffer size check}'') that goes beyond the literal syntax changes in the diff, from which the intent can be too difficult to infer.
This discrepancy between the commit message and code diff makes having both sources of information meaningful, and removing the diff (\textsf{Only msg}) or the message (\textsf{Only diff}) causes a performance drop from \review{1.27} to \review{1.08} and \review{1.05} (meaning `at least reaching'), respectively.
}
We seek deeper case studies on the informativeness of commit messages and its impact on test generation in \textbf{RQ3}.

\textbf{Task Difficulty}.
We evaluate the impact of having \toolname generate the command-line prompt in addition to the regression test case (\textsf{Gen. cmd}) for \libxml, \poppler, \review{and \jq}, which require a specific command-line prompt to find or reproduce the bug. Our results show \review{a noticeable but limited negative effect on \toolname's performance for 7 out of 28 (25\%) commits,} changing the overall average score from \review{1.37} to \review{1.19} and \review{1.27} to \review{1.08} for the bug-finding and bug-reproduction scenarios, respectively. \review{This is primarily because some programs have vast input space for command line prompts, while the bugs can only be triggered by specific values. For example, \jq~\#3262 was triggered by command \texttt{jq .[54E100]=7} in the original PoC. \toolname could reliably generate commands with the array index feature to reach code changes, but only reproduced the bug in 2 out of 10 repetitions.}
This suggests that the LLM is capable of generating \review{relevant} command-line prompts for regression testing, \review{though it struggles when the input space is vast}.

Interestingly, \toolname found one otherwise \emph{undiscovered bug} when asked to also generate the command line prompt for reproducing Issue~\#550 of \libxml2. This bug was unrelated to Issue~\#550 but the bug-triggering command line {\small\texttt{-{}-xpath -{}-dropdtd}} was related to the \emph{same feature} that was changed.
This confirms our intuition that \toolname generates regression test cases from the intention more than the actual code changes.

\textbf{LLM Module}.
We evaluate the impact of LLM size on \toolname's performance and find that model power significantly affects effectiveness.
Using a weaker LLM (\gptmini) reduces effectiveness for \review{16} issues in the bug-finding scenario, with test cases failing to reach code changes in \review{5} BICs. The average score drops from \review{1.37} to \review{0.97}.
For the bug reproduction scenario, effectiveness was reduced for \review{12} patches. In \review{4} cases, test cases fail to reach code changes, and the average score drops from \review{1.27} to \review{1.01}.
Conversely, a stronger LLM (\deepseekr) enables \toolname to find more bugs.
In the bug-finding scenario, the number of bugs increased from \review{7 to 10}, with 1 fewer in \jerryscript, but 4 additional bugs in \jq, \poppler, \php, \micropython, respectively, while \gpt failed to find any for the last three subjects. In the bug-reproduction scenario, the number of detected bugs increases from \review{13} to \review{20}, with 1 fewer in \mujs, but 2 additional bugs in \jerryscript, \review{2 in \jq, and 4 in \micropython}. The average score rises from \review{1.37} to \review{1.62} in the bug-finding scenario and from \review{1.27} to \review{1.62} in the bug-reproduction scenario.
These results indicate that effectiveness heavily depends on the LLM's size/capacity, particularly for generating complex inputs.

We also evaluate an increase in temperature (i.e., the degree of confabulation). The score increased from \review{1.37} to \review{1.43} for bug-finding and \review{1.27} to \review{1.28} for bug-reproduction. This implies that maximizing temperature would also lead to an increase in test case diversity.

\textbf{Number of feedback iterations}.
We evaluate the impact on effectiveness if we increase the number of feedback iterations from five (5) to ten (10) and find small positive impact for bug-finding (from \review{1.37} to \review{1.43}) and bug-reproduction (from \review{1.27} to \review{1.33}).
In every iteration, \toolname appends the execution feedback of the previous iteration from the Execution Analyser to the synthesized prompt (Fig.~\ref{fig:prompt0}).
After increasing the number of iterations,
  one case (1) turned from failing to reaching;
  two cases (2) turned from reaching to bug-triggering;
  one case (1) turned from output-changing to bug-triggering;
Overall, we see the greatest change into the bug-triggering category when number of iterations is increased, at the cost of execution time.

\textbf{Feedback utility}.
We evaluate the impact of dropping the execution analysis result from the feedback, i.e., having only the record of previously generated inputs in the prompt, and find that it has a negative impact on the effectiveness of \toolname.
In the bug-finding scenario (from \review{1.37} to \review{1.19}), the decrease is obvious for $\langle$\poppler, \#1289$\rangle$, $\langle$\php, \#17463$\rangle$ \review{, and $\langle$\micropython, \#13007$\rangle$}, where the generated test case now even fails to reach the code changes.
In the bug-reproduction scenario (from \review{1.27} to \review{1.21}), \toolname's effectiveness reduces especially for $\langle$\poppler, \#1303$\rangle$ and \review{$\langle$\micropython, \#13007$\rangle$} where the generated test case now fails to reach the code changes.
The result indicates that the execution feedback is crucial for generating the regression test input.

Taking a closer look, we found that the execution feedback was especially helpful to the LLM by pointing out invalid components in the input. For instance, the test case generated for \mujs in the first iteration often contains token {\small\texttt{console}}, which is unsupported by the program and causes error: {\small\texttt{ReferenceError: `console' is not defined}}. The feedback enables the LLM to refine its inputs, progressively avoiding invalid components until it finally generates a valid, bug-triggering input.

\result{\textbf{RQ2}.
  We found that \toolname continues to generate effective test cases that find bugs in the changed feature as long as the intention is obvious which feature is changed, i.e., even when only an expressive commit message is provided.
  We also found that asking \toolname to generate the command line prompt itself had only a limited negative effect on its performance; it even found a bug that was related to the changed feature but was only fixed much later.
  Reducing the model size or dropping the execution feedback both reduced the effectiveness.
  Doubling the number of iterations (at double the cost) increased the effectiveness of \toolname.
}

\subsection{RQ3. Effectiveness of Commit Messages}

To investigate how informative and self-contained commit messages need to be for generating regression test cases without access to the code, we enhance or reduce the information in commit messages and observe how \toolname's performance changes under these modifications, compared to using the original commit messages alone. In this study, we focus exclusively on bug reproduction because the intention behind a bug-fixing change is straightforward, while bug-inducing commits often conflate multiple intentions (e.g., adding a feature and refactoring), which makes them less suitable for controlled evaluation.

\begin{figure}[t]
\newcommand{\tikzsymbol}[2][circle]{\tikz[baseline=-0.5ex]\node[inner
sep=2pt,shape=#1,draw,#2]{};}%
\begin{minipage}[t]{0.35\textwidth}
  \centering
  \includegraphics[width=.9\textwidth, trim=0 0 0 0,clip]{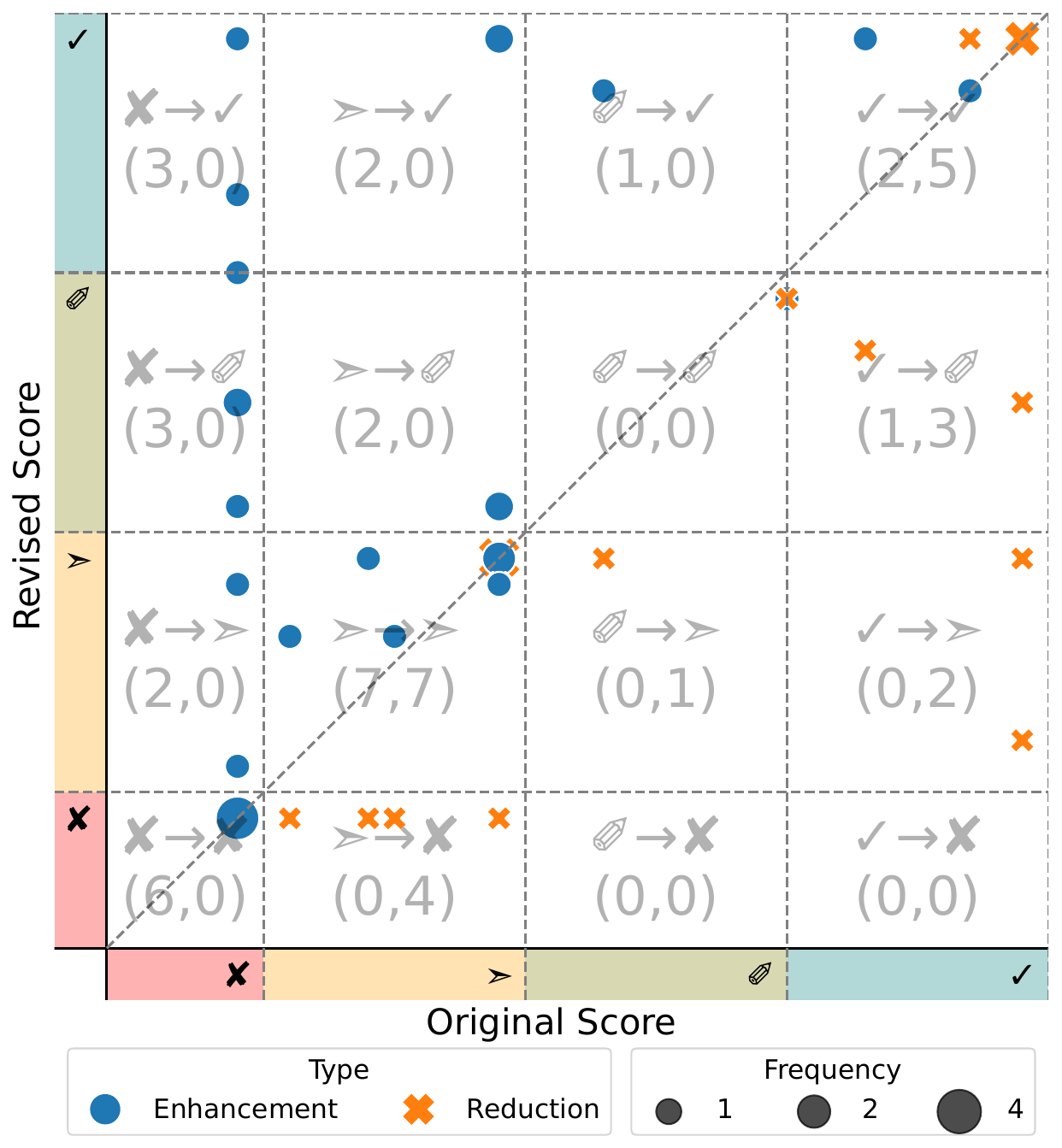}
  \captionsetup{skip=2pt}
  \captionof{figure}{
    Changes in bug reproduction scores from original (x-axis) to modified (y-axis) commit messages. \tikzsymbol{fill=blue} and \textcolor{orange}{\ding{54}} mark enhancements and reductions (higher is better).
    The background label $(X, Y)$ means that $X$ and $Y$ scores moved here after enhancement and reduction, respectively. 
    }
  \label{fig:rq3}
\end{minipage}%
\hfill
\begin{minipage}[t]{0.63\textwidth}
  \vspace*{-13.5em}
  \captionof{table}{Examples of how intention in commit message affects test case generation.
  (Enhancement: {\color{teal}\textit{teal \& italic}}, Reduction: {\color{red}\textbf{red \& bold}})
  }
  \label{tab:rq3_examples}
  \resizebox{\textwidth}{!}{%
    \begin{tabular}{c|l|c|l|c}
      \toprule
      \textbf{(Subject, Issue)}               & \multicolumn{1}{|c|}{\textbf{Message}}                                                   & \textbf{Ver.}                & \multicolumn{1}{|c|}{\textbf{Generated Input}} & \textbf{Result} \\ \midrule
      \multirow{7}{*}{(\zthree, \#7252)}      & \multirow{7}{*}{\makecell[l]{fix \#7252 \textcolor{teal}{\textit{segmentation fault in}}                                                                                                   \\\textcolor{teal}{\textit{nested re.loop formula processing}}\\\textcolor{teal}{\textit{when handling specific boundary}}                                                                                             \\{\color{teal}\textit{values for nested re.loop}}\\\textcolor{teal}{\textit{expressions with same bounds of}}\\\textcolor{teal}{\textit{0 and 1}}}} & \multirow{3}{*}{Orig.}
                                              & \multirow{3}{*}{\texttt{\small \makecell[l]{ (declare-fun a () Int)                                                                                                                        \\(assert (= a (+ a 1)))\\(check-sat)}}}                                                                                         & \multirow{3}{*}{\huge\xmark}                                                     \\
                                              &                                                                                          &                              &                                                &                 \\
                                              &                                                                                          &                              &                                                &                 \\ \cline{3-5}
                                              &                                                                                          & \multirow{4}{*}{Enhc.}
                                              & \multirow{4}{*}{\texttt{\small \makecell[l]{ (declare-fun a () (RegEx String))                                                                                                             \\ (assert (str.in.re "a" (\\\ \ re.loop (re.loop a 0 1) 0 1)))\\(check-sat)}}}                  & \multirow{4}{*}{\huge\cmark}                                                                                                                                                                                                           \\
                                              &                                                                                          &                              &                                                &                 \\
                                              &                                                                                          &                              &                                                &                 \\
                                              &                                                                                          &                              &                                                &                 \\ \midrule
      \multirow{6}{*}{(\mujs, \#65)}          & \multirow{6}{*}{\makecell[l]{Fix issue \#65: Uninitialized name                                                                                                                            \\\textcolor{teal}{\textit{and length properties}} \\in Function.prototype \textcolor{teal}{\textit{}}                                                              \\\textcolor{teal}{\textit{causing a null pointer dereference}}\\\textcolor{teal}{\textit{when `Function.prototype` is}}\\\textcolor{teal}{\textit{converted to a string}}}} & \multirow{3}{*}{Orig.}
                                              & \multirow{3}{*}{\texttt{\small \makecell[l]{ Function.prototype.call.name;}}}            & \multirow{3}{*}{\huge\rmark}                                                                    \\
                                              &                                                                                          &                              &                                                &                 \\
                                              &                                                                                          &                              &                                                &                 \\ \cline{3-5}
                                              &                                                                                          & \multirow{3}{*}{Enhc.}
                                              & \multirow{3}{*}{\texttt{\small \makecell[l]{ String(Function.prototype);}}}              & \multirow{3}{*}{\huge\cmark}                                                                    \\
                                              &                                                                                          &                              &                                                &                 \\
                                              &                                                                                          &                              &                                                &                 \\ \midrule
      \multirow{6}{*}{\makecell[c]{(\jerryscript,\\\#5153)}} & \multirow{6}{*}{\makecell[l]{Fix parsing of invalid asterisk                                                                                                                               \\symbol after private field symbol\\\textcolor{teal}{\textit{, which previously caused a}}                                                                                          \\\textcolor{teal}{\textit{segmentation fault when an}}\\\textcolor{teal}{\textit{asterisk was used after an async}}\\\textcolor{teal}{\textit{private method declaration.}}}} & \multirow{3}{*}{Orig.}
                                              & \multirow{3}{*}{\texttt{\small \makecell[l]{ class A \{ \#private* \}}}}                 & \multirow{3}{*}{\huge\xmark}                                                                    \\
                                              &                                                                                          &                              &                                                &                 \\
                                              &                                                                                          &                              &                                                &                 \\\cline{3-5}
                                              &                                                                                          & \multirow{3}{*}{Enhc.}
                                              & \multirow{3}{*}{\texttt{\small \makecell[l]{ class C \{ async \#*foo() \{\} \}}}}        & \multirow{3}{*}{\huge\cmark}                                                                    \\
                                              &                                                                                          &                              &                                                &                 \\
                                              &                                                                                          &                              &                                                &                 \\ \midrule
      \multirow{8}{*}{\makecell[c]{(\jerryscript,\\\#5117)}} & \multirow{8}{*}{\makecell[l]{Fix arguments object in                                                                                                                                       \\\textcolor{red}{\textbf{eval-ed}} functions in\\static class initializers}}                                                                & \multirow{4}{*}{Orig.}
                                              & \multirow{4}{*}{\texttt{\small \makecell[l]{ class MyClass \{ static \{                                                                                                                    \\\ \ eval("function testFunc() \{\\\ \ \ \ \ \ \ \ \ \ return arguments; \}");\\\ \ console.log(testFunc(1, 2, 3));\}\}}}}           & \multirow{4}{*}{\huge\cmark}                                                     \\
                                              &                                                                                          &                              &                                                &                 \\
                                              &                                                                                          &                              &                                                &                 \\
                                              &                                                                                          &                              &                                                &                 \\\cline{3-5}
                                              &                                                                                          & \multirow{4}{*}{Redu.}
                                              & \multirow{4}{*}{\texttt{\small \makecell[l]{ class MyClass \{ static \{                                                                                                                    \\\ \ function testFunc() \{                                                                                                                                                                \\\ \ \ \ console.log(arguments); \}\\\ \ testFunc(1, 2, 3);\}\}}}}                           & \multirow{4}{*}{\huge\rmark}                                                     \\
                                              &                                                                                          &                              &                                                &                 \\
                                              &                                                                                          &                              &                                                &                 \\
                                              &                                                                                          &                              &                                                &                 \\
      \bottomrule
    \end{tabular}}
\end{minipage}
\vspace{-2em}
\end{figure}

\noindent
Specifically, we conduct the following procedure:
\begin{enumerate}[leftmargin=*,itemsep=0.05cm,topsep=0.1cm]
\item We first examine each of the \review{36} BFC and its corresponding bug report to identify the key elements describing the bug and its resolution.
\item \emph{Commit Message Reduction.} For BFCs where \toolname was able to generate test cases that at least reach the modified code (i.e., \review{22} BFCs), we manually remove some of the identified key elements from the commit message, leaving a reduced version with less critical information.
For a representative example, see last row in Table~\ref{tab:rq3_examples}.
\item \emph{Commit Message Enhancement.} There are \review{7} BFCs where \toolname always generates a bug-triggering test case. For all other BFCs (i.e., \review{29} BFCs), we enhance the commit message. To standardize the enhancement process across commits, we employ another LLM (\texttt{gemini-2.5-flash}) which is given the original commit and the corresponding bug report. We then manually verify that the enhanced message includes the key elements identified in Step~1. The details of the LLM configuration and prompts are provided in the Supplementary Material \cite{supplementary}. For representative examples of such enhancements, see top three rows in Table~\ref{tab:rq3_examples}.%
\end{enumerate}

\autoref{fig:rq3} summarizes the changes in \toolname's effectiveness when using modified commit messages compared to the original ones. The x-axis and y-axis denote effectiveness scores with the original and modified commit messages, respectively. Circles and crosses mark enhancements and reductions, with marker size showing commit counts. Markers above the diagonal represent improvements, those below represent decreases, and the diagonal line indicates no change.

The results show that enhancing commit messages generally improves \toolname's effectiveness. This suggests that well-detailed commit messages substantially help the LLM capture the intention of the change and generate more effective test cases. After enhancement, \review{59\% (17/29)} of the commits exhibited improved bug reproduction scores. Notably, among \review{14} commits where \toolname initially failed to generate test cases that reached the modified code, \review{5} were able to trigger the bug, while 2 and 1 managed to change the output and reach the modified code, respectively, after enhancement. Combining these with the cases where the original commit message already triggered the bug, \toolname successfully generated bug-triggering tests in \review{18} out of \review{36} cases, given only the commit message-level natural language description of the code change.

To gain insights into the results, we manually examined the quality of the original commit messages to understand why they sometimes failed to guide regression test generation, and how the enhanced messages improved the outcome.

In failing cases (either \xmark or \rmark), we found that some original messages contained almost no useful information (e.g. \zthree's ``\textsf{Fix \#IssueId}''), leaving \toolname with little basis for generating test cases that could even reach the modified code. Others described only the root cause in the implementation without mentioning the bug's manifestation (e.g., (\mujs, \#65)'s ``\textsf{Fix: Uninitialized name in Function.prototype function}''), which made it difficult for \toolname to infer the developer’s intention and produce effective tests. Most commit messages did reference the fix, but often lacked precision in describing the conditions under which the bug occurred (e.g. (\jerryscript, \#5155)'s ``\textsf{Fix parsing of invalid asterix symbol after private field symbol}'').
After enhancement—when the commit messages included details about the bug’s nature and manifestation (\autoref{tab:rq3_examples})—\toolname was able to infer the bug-triggering context and generate effective bug-revealing test cases.

We observed only \review{two} cases where the score dropped slightly (from 1.0 to 0.9 and 2.1 to 2.0, respectively).
On manual inspection, we found that, after enhancement, the message led \toolname to generate malformed PDF inputs that, in one of ten runs, triggered parser syntax errors before reaching the modified code. This drop stems from execution randomness rather than a weakness of the enhancement.

Conversely, reducing commit messages tends to decrease \toolname's effectiveness. This not only reinforces our claim but also confirms that the finding of \textbf{RQ2}—that commit messages alone are sufficient to generate effective test cases—is not coincidental or merely a result of LLM memorization.
After reduction, \review{50\% (11/22)} of the commits exhibited decreased bug reproduction scores, with a \review{median} of only two words reduced. Notably, among \review{13} commits where \toolname initially generated test cases that triggered bug, \review{1 could not even reach modified code}, and \review{4} were only able to reach the code. Among \review{9} commits where \toolname initially generated test cases that could reach the changed code, \review{3} were not even able to reach the code after reduction.
For example, by removing just one word ``eval-ed'' from the commit message of \jerryscript \#5117, the generated tests can only reach the changed code but never trigger the bug.

\result{\textbf{RQ3}.
Our results demonstrate that \toolname's ability to generate effective regression test cases from commit messages alone is not coincidental: in \review{11/22} cases, the effectiveness score dropped when key information about the bug was removed from the commit message. Furthermore, when sufficient detail is present, \toolname can produce effective regression test cases. With enhanced commit messages, \review{18/36} cases were able to trigger the bug using only the natural language description in the commit message. This result highlights that LLMs are capable of generating regression tests solely from natural language descriptions of code changes.
}

\subsection{RQ4. Comparison to the State-of-the-Art}

\begin{table}[!htbp]
  \caption{
    Results for bug finding and bug reproduction effectiveness of \toolname, \toolname + fuzzing (\toolnamefuzz), and WAFLGo across ten repetitions on the \review{50} subjects where WAFLGo was runnable. For each configuration, we report initial seed effectiveness (\xmark to \cmark), number of WAFLGo campaigns that found the bug and time-to-exposure (\emph{T.O.} = timeout after 24 hours), \toolname's average effectiveness score (slider), number of commit-reaching / output-changing / bug-triggering repetitions, and time spent. \review{`\texttt{-/-}' indicates \toolnamefuzz was skipped as \toolname already caught the bug in 10/10 repetitions. [Bug\textsubscript{all}] is union of them.}\vspace{-0.2cm}}
  \label{tab:clevfuzz}

  \resizebox{0.99\textwidth}{!}{%
    \begin{tabular}{c|cc|crr|crrrr|rrr}
      \toprule
      \multirow{2}{*}{}
                                                                       & \multirow{2}{*}{\begin{tabular}{@{}c@{}}Subject\\(\# seeds)\end{tabular}} & \multirow{2}{*}{Issue} & \multicolumn{3}{c|}{WAFLGo} & \multicolumn{5}{c|}{\toolname} & \multicolumn{3}{c}{\toolnamefuzz}                                                                                                                                                               \\
                                                                       &                                                                           &                        & Init                        & Bug                            & T (h:m:s)                         & \xmark~\rmark~\dmark~\cmark  & Reach           & Change         & Bug             & T (h:m:s) & Bug               & T (h:m:s)      & Bug\textsubscript{all} \\ \midrule
      \multirow{27}{*}{\begin{sideways}Bug-finding\end{sideways}} & \multirow{4}{*}{\makecell{\mujs\\(19)}} & \#65 & \xmark & 10/10 & 04:43:56 & \SLIDER{3.3em}{1.00}{circle} & 10/10 & 10/10 & 10/10 & 00:00:07 & -/- & - & 10/10 \\
      &  & \#141 & \xmark & 0/10 & T.O. & \SLIDER{3.3em}{0.67}{circle} & 10/10 & 10/10 & 0/10 & 00:00:45 & 10/10 & 03:25:05 & 10/10 \\
      &  & \#145 & \xmark & 10/10 & 00:17:21 & \SLIDER{3.3em}{0.87}{circle} & 10/10 & 8/10 & 8/10 & 00:00:37 & 2/2 & 00:00:00 & 10/10 \\
      &  & \#166 & \cmark & 10/10 & T.O. & \SLIDER{3.3em}{0.37}{circle} & 10/10 & 1/10 & 0/10 & 00:00:57 & 0/10 & T.O. & 0/10 \\
      \cmidrule{2-14}
      & \multirow{5}{*}{\makecell{\libxml\\(14)}} & \#535 & \cmark & 10/10 & T.O. & \SLIDER{3.3em}{1.00}{circle} & 10/10 & 10/10 & 10/10 & 00:00:05 & -/- & - & 10/10 \\
      &  & \#550 & \dmark & 0/10 & T.O. & \SLIDER{3.3em}{0.40}{circle} & 10/10 & 2/10 & 0/10 & 00:00:27 & 0/10 & T.O. & 0/10 \\
      &  & \#553 & \cmark & 10/10 & T.O. & \SLIDER{3.3em}{1.00}{circle} & 10/10 & 10/10 & 10/10 & 00:00:06 & -/- & - & 10/10 \\
      &  & \#720 & \dmark & 0/10 & T.O. & \SLIDER{3.3em}{0.33}{circle} & 10/10 & 0/10 & 0/10 & 00:00:53 & 4/10 & 18:08:01 & 4/10 \\
      &  & \#841 & \cmark & 10/10 & T.O. & \SLIDER{3.3em}{1.00}{circle} & 10/10 & 10/10 & 10/10 & 00:00:05 & -/- & - & 10/10 \\
      \cmidrule{2-14}
      & \multirow{5}{*}{\makecell{\poppler\\(100)}} & \#1282 & \rmark & 6/10 & 00:06:23 & \SLIDER{3.3em}{0.37}{circle} & 8/10 & 3/10 & 0/10 & 00:03:16 & 8/8 & 00:00:10 & 8/10 \\
      &  & \#1289 & \rmark & 0/10 & T.O. & \SLIDER{3.3em}{0.13}{circle} & 4/10 & 0/10 & 0/10 & 00:02:45 & 3/4 & 06:00:04 & 3/10 \\
      &  & \#1303 & \xmark & 0/10 & T.O. & \SLIDER{3.3em}{0.00}{circle} & 0/10 & 0/10 & 0/10 & 00:03:38 & -/- & - & 0/10 \\
      &  & \#1305 & \xmark & 0/10 & T.O. & \SLIDER{3.3em}{0.00}{circle} & 0/10 & 0/10 & 0/10 & 00:04:27 & -/- & - & 0/10 \\
      &  & \#1381 & \rmark & 0/10 & T.O. & \SLIDER{3.3em}{0.00}{circle} & 0/10 & 0/10 & 0/10 & 00:03:22 & -/- & - & 0/10 \\
      \cmidrule{2-14}
      & \multirow{3}{*}{\makecell{\jerryscript\\(19)}} & \#5117 & \rmark & 4/10 & 00:13:02 & \SLIDER{3.3em}{0.67}{circle} & 10/10 & 10/10 & 0/10 & 00:01:11 & 10/10 & 00:01:41 & 10/10 \\
      &  & \#5138 & \rmark & 0/10 & T.O. & \SLIDER{3.3em}{0.73}{circle} & 10/10 & 10/10 & 2/10 & 00:01:18 & 8/8 & 00:00:37 & 10/10 \\
      &  & \#5153 & \rmark & 0/10 & T.O. & \SLIDER{3.3em}{0.67}{circle} & 10/10 & 10/10 & 0/10 & 00:01:35 & 10/10 & 00:00:59 & 10/10 \\
      \cmidrule{2-14}
      & \multirow{4}{*}{\makecell{\jq\\(38)}} & \#2825 & \rmark & 0/10 & T.O. & \SLIDER{3.3em}{0.33}{circle} & 10/10 & 0/10 & 0/10 & 00:00:23 & 10/10 & 04:42:34 & 10/10 \\
      &  & \#2976 & \rmark & 2/10 & 01:55:03 & \SLIDER{3.3em}{0.33}{circle} & 10/10 & 0/10 & 0/10 & 00:00:22 & 10/10 & 01:21:39 & 10/10 \\
      &  & \#3262 & \cmark & 10/10 & T.O. & \SLIDER{3.3em}{1.00}{circle} & 10/10 & 10/10 & 10/10 & 00:00:09 & -/- & - & 10/10 \\
      &  & \#49014 & \xmark & 0/10 & T.O. & \SLIDER{3.3em}{0.13}{circle} & 2/10 & 2/10 & 0/10 & 00:00:32 & 2/2 & 07:51:24 & 2/10 \\
      \cmidrule{2-14}
      & \multirow{5}{*}{\makecell{\micropython\\(13)}} & \#13007 & \rmark & 0/10 & T.O. & \SLIDER{3.3em}{0.53}{circle} & 9/10 & 7/10 & 0/10 & 00:00:27 & 7/9 & 07:07:12 & 7/10 \\
      &  & \#13041 & \xmark & 0/10 & T.O. & \SLIDER{3.3em}{0.00}{circle} & 0/10 & 0/10 & 0/10 & 00:00:31 & -/- & - & 0/10 \\
      &  & \#17733 & \rmark & 0/10 & T.O. & \SLIDER{3.3em}{0.33}{circle} & 10/10 & 0/10 & 0/10 & 00:00:26 & 6/10 & 12:35:08 & 6/10 \\
      &  & \#17815 & \rmark & 0/10 & T.O. & \SLIDER{3.3em}{0.63}{circle} & 10/10 & 9/10 & 0/10 & 00:00:26 & 4/10 & 17:54:34 & 4/10 \\
      &  & \#17847 & \xmark & 0/10 & T.O. & \SLIDER{3.3em}{0.67}{circle} & 10/10 & 10/10 & 0/10 & 00:00:40 & 2/10 & 20:33:41 & 2/10 \\
      \cmidrule{2-14}
      & \multicolumn{2}{c|}{Aggregate} &  & \textbf{10/26} & \textbf{19:39:50} & \SLIDER{3.3em}{0.51}{circle} & \textbf{22/26} & \textbf{17/26} & \textbf{7/26} & \textbf{00:01:08} &  & \textbf{05:42:01} & \textbf{20/26} \\
      \midrule
      \multirow{25}{*}{\begin{sideways}Bug-reproduction\end{sideways}} & \multirow{4}{*}{\makecell{\mujs\\(19)}} & \#65 & \rmark & 0/10 & T.O. & \SLIDER{3.3em}{0.73}{circle} & 10/10 & 6/10 & 6/10 & 00:00:19 & 1/4 & 18:03:42 & 7/10 \\
      &  & \#141 & \xmark & 5/10 & 00:12:47 & \SLIDER{3.3em}{0.70}{circle} & 10/10 & 10/10 & 1/10 & 00:00:31 & 9/9 & 01:43:32 & 10/10 \\
      &  & \#145 & \xmark & 10/10 & 00:08:26 & \SLIDER{3.3em}{0.93}{circle} & 10/10 & 9/10 & 9/10 & 00:00:28 & 1/1 & 00:00:03 & 10/10 \\
      &  & \#166 & \cmark & 10/10 & T.O. & \SLIDER{3.3em}{1.00}{circle} & 10/10 & 10/10 & 10/10 & 00:00:07 & -/- & - & 10/10 \\
      \cmidrule{2-14}
      & \multirow{5}{*}{\makecell{\libxml\\(14)}} & \#535 & \cmark & 10/10 & T.O. & \SLIDER{3.3em}{1.00}{circle} & 10/10 & 10/10 & 10/10 & 00:00:05 & -/- & - & 10/10 \\
      &  & \#550 & \xmark & 0/10 & T.O. & \SLIDER{3.3em}{0.00}{circle} & 0/10 & 0/10 & 0/10 & 00:00:29 & -/- & - & 0/10 \\
      &  & \#553 & \cmark & 10/10 & T.O. & \SLIDER{3.3em}{1.00}{circle} & 10/10 & 10/10 & 10/10 & 00:00:13 & -/- & - & 10/10 \\
      &  & \#720 & \dmark & 0/10 & T.O. & \SLIDER{3.3em}{0.23}{circle} & 7/10 & 0/10 & 0/10 & 00:01:44 & 7/7 & 03:09:11 & 7/10 \\
      &  & \#841 & \cmark & 10/10 & T.O. & \SLIDER{3.3em}{1.00}{circle} & 10/10 & 10/10 & 10/10 & 00:00:05 & -/- & - & 10/10 \\
      \cmidrule{2-14}
      & \multirow{5}{*}{\makecell{\poppler\\(100)}} & \#1282 & \rmark & 0/10 & T.O. & \SLIDER{3.3em}{0.30}{circle} & 9/10 & 0/10 & 0/10 & 00:02:03 & 1/9 & 22:09:30 & 1/10 \\
      &  & \#1289 & \rmark & 0/10 & T.O. & \SLIDER{3.3em}{0.67}{circle} & 8/10 & 6/10 & 6/10 & 00:01:30 & 2/2 & 00:00:08 & 8/10 \\
      &  & \#1303 & \xmark & 0/10 & T.O. & \SLIDER{3.3em}{0.07}{circle} & 1/10 & 1/10 & 0/10 & 00:02:12 & 0/1 & T.O. & 0/10 \\
      &  & \#1305 & \rmark & 0/10 & T.O. & \SLIDER{3.3em}{0.00}{circle} & 0/10 & 0/10 & 0/10 & 00:02:27 & -/- & - & 0/10 \\
      &  & \#1381 & \rmark & 0/10 & T.O. & \SLIDER{3.3em}{0.00}{circle} & 0/10 & 0/10 & 0/10 & 00:02:27 & -/- & - & 0/10 \\
      \cmidrule{2-14}
      & \multirow{3}{*}{\makecell{\jerryscript\\(19)}} & \#5117 & \rmark & 0/10 & T.O. & \SLIDER{3.3em}{0.67}{circle} & 10/10 & 5/10 & 5/10 & 00:00:50 & 2/5 & 20:54:43 & 7/10 \\
      &  & \#5138 & \xmark & 0/10 & T.O. & \SLIDER{3.3em}{0.33}{circle} & 10/10 & 0/10 & 0/10 & 00:01:14 & 4/10 & 17:50:21 & 4/10 \\
      &  & \#5013 & \xmark & 0/10 & T.O. & \SLIDER{3.3em}{0.33}{circle} & 10/10 & 0/10 & 0/10 & 00:01:12 & 10/10 & 03:02:42 & 10/10 \\
      \cmidrule{2-14}
      & \multirow{4}{*}{\makecell{\jq\\(38)}} & \#2825 & \rmark & 0/10 & T.O. & \SLIDER{3.3em}{0.37}{circle} & 10/10 & 1/10 & 0/10 & 00:00:28 & 2/5 & 19:20:31 & 2/10 \\
      &  & \#2976 & \rmark & 1/10 & 19:03:38 & \SLIDER{3.3em}{1.00}{circle} & 10/10 & 10/10 & 10/10 & 00:00:03 & -/- & - & 10/10 \\
      &  & \#3262 & \cmark & 10/10 & T.O. & \SLIDER{3.3em}{1.00}{circle} & 10/10 & 10/10 & 10/10 & 00:00:08 & -/- & - & 10/10 \\
      &  & \#49014 & \xmark & 0/10 & T.O. & \SLIDER{3.3em}{0.00}{circle} & 0/10 & 0/10 & 0/10 & 00:00:32 & -/- & - & 0/10 \\
      \cmidrule{2-14}
      & \multirow{3}{*}{\makecell{\micropython\\(13)}} & \#13041 & \dmark & 0/10 & T.O. & \SLIDER{3.3em}{0.67}{circle} & 10/10 & 10/10 & 0/10 & 00:00:21 & 10/10 & 02:58:33 & 10/10 \\
      &  & \#17733 & \dmark & 0/10 & T.O. & \SLIDER{3.3em}{0.33}{circle} & 10/10 & 0/10 & 0/10 & 00:00:31 & 5/10 & 15:00:18 & 5/10 \\
      &  & \#17847 & \dmark & 0/10 & T.O. & \SLIDER{3.3em}{0.00}{circle} & 0/10 & 0/10 & 0/10 & 00:00:29 & -/- & - & 0/10 \\
      \cmidrule{2-14}
      & \multicolumn{2}{c|}{Aggregate} &  & \textbf{8/24} & \textbf{21:48:32} & \SLIDER{3.3em}{0.51}{circle} & \textbf{19/24} & \textbf{14/24} & \textbf{11/24} & \textbf{00:00:51} &  &  \textbf{06:11:24} & \textbf{18/24} \\
      \bottomrule
    \end{tabular}}
  \vspace{-0.2cm}
\end{table}

In this section, we compare \toolname against two baselines: primarily WAFLGo, the state-of-the-art in regression test generation, and existing LLM-based test generation approaches~\cite{Fuzz4ALL}. Since prior LLM-based techniques are not designed for directed, regression-focused testing, we instead compare against unguided LLM-based test generation with no commit information used. We briefly report this comparison at the end of the section, with full details in the Supplementary Material \cite{supplementary}.

For WAFLGo, we found that it does not work for any commits of Z3 and PHP due to its static analysis module exceeding our server's 251GB RAM. It also failed to build for commits \texttt{b7e3bae} and \texttt{de51531} of \jerryscript and \review{commits \texttt{e702046}, \texttt{f397a3e}, \texttt{c0252d7}, and \texttt{4b013ec} of \micropython}. Hence, we have \review{50} commits for the comparison. \review{Finally, for WAFLGo, we report only the fuzzing campaign time, as done in prior work, and do not include the static analysis pre-processing overhead required to construct its directed fuzzing guidance---a choice that favors WAFLGo in our comparison.}

\subsubsection{RQ4.1\quad Comparison with WAFLGo}

\autoref{tab:clevfuzz} contains the results for the comparison. We first evaluate both tools head to head and then explore the opportunities to fuzz interesting \toolname-generated test cases (\toolnamefuzz).

\textbf{Effectiveness}.
In the bug reproduction scenario, \toolname (without initial seeds) outperforms WAFLGo (which requires seeds): it reproduces bugs in \review{11 of 24} BFCs, compared to WAFLGo’s \review{8 of 24}. This is a notable improvement, especially given that \toolname operates without a seed corpus.
In the bug finding scenario, \toolname reaches the modified code in \review{22 of 26} BICs and changes the output in \review{17 of them}, both higher than WAFLGo. However, it triggers the actual bug in \review{7 of 26} BICs, fewer than WAFLGo (\review{10 of 26}). This result is expected, since \toolname relies on the intention expressed in commit messages (as shown in \textbf{RQ3}), and BICs often describe new features or improvements rather than the bug itself.
We further examine WAFLGo's dependence on the initial seed corpus in the next section.

\textbf{Execution time}.
\toolname is substantially faster than WAFLGo, making our LLM-based approach more suitable as part of the CI/CD pipeline, which runs under strict time and resource constraints. A typical fuzzing campaign is set to 24 hours. Apart from ten cases where the initial seeds already exposed the bug introduced or patched by a commit, WAFLGo takes between five minutes and five hours on average to find the bugs, and otherwise times out.
For the commit fixing Issue~\#2976 of \jq, WAFLGo took more than 19 hours. \toolname's execution time is bounded by the number of feedback iterations (fixed to 5 in our experiments) and requires \icse{roughly a minute for \mujs, \libxml, \jerryscript, \jq, and \micropython and mostly less than three minutes for \poppler, \zthree, and \php.}

\textbf{Fuzzing \toolname's seeds for improved effectiveness}.
Since fuzzing works well if seeds are close already, why not try using the \toolname-generated inputs as fuzzer seeds (\toolnamefuzz)?
During qualitative analysis of change-reaching and output-changing \toolname test cases for \textbf{RQ1} (\S\ref{sec:rq1}), we found that they are often not ``very far'' from bug-revealing; a few modification can turn them to bug-finding test cases. This idea is further inspired by LLMs being used as seed generators for fuzzing, a successful strategy in the 2024 DARPA AI Cyberchallenge (AIxCC).\footnote{Trail of Bits, one of the finalists says: ``Our system uses large language models (LLMs) to generate seed inputs for fuzzing, significantly reducing the time needed to discover vulnerabilities. This innovative approach helps us work within the competition's strict time constraints.''~\cite{TrailBitsButtercup2024}}

For each trial with a change-reaching or output-changing \toolname-generated input, we run a 24-hour AFL++ v4.21c campaign with the default configuration, using that test case as the only seed. The denominators in Column~[\toolnamefuzz. Bug] of \autoref{tab:clevfuzz} indicate the number of such trials among \review{50} subjects. Specifically, \review{143} \toolname-generated test cases reached the changes but failed to reveal the bug in \review{17} BICs, while \review{83} reached the change but did not reveal the bug in \review{13} BFCs. The full results across all \review{72} commits are provided in Supplementary Material \cite{supplementary}.

\textbf{Results}. As shown in the last column of~\autoref{tab:clevfuzz}, our LLM-seeded fuzzer \toolnamefuzz outperforms WAFLGo, which is started on a user-provided seed corpus. Overall, \toolnamefuzz finds bugs in \review{20 out of 26} BICs and \review{18 out of 24} BFCs, \review{significantly} outperforming WAFLGo (\review{10 and 8}), \review{finding more than twice as many bugs} in both scenarios.

Specifically, \toolnamefuzz can find \review{13} additional bugs for BICs and \review{7} for BFCs after the fuzzing campaign. The fuzzing turned \review{96 out of 143 (67\%)} \toolname-generated test cases for BICs and \review{54 out of 83 (65\%)} for BFCs into bug-revealing test cases. This indicates that \toolname-generated test cases are often close to bug-revealing already, and fuzzing can effectively explore the neighborhood of these test cases to find the bugs.

Timewise, this LLM-based zero-shot fuzzing approach remains more than twice as fast as WAFLGo. Considering the combined time for \toolname and fuzzing, \toolnamefuzz takes roughly \review{6} hours\footnote{Including the time-out of the fuzzing campaign, which is 24 hours.}  for the subjects where WAFLGo is applicable, while WAFLGo takes roughly \review{20 and 22} hours for bug finding and bug reproduction scenarios, respectively.
Indeed, this is an intriguing result as WAFLGo can be considered few-shot (i.e., starting from a user-provided corpus) while our approach is zero-shot (i.e., no examples needed).

\subsubsection{RQ4.2\quad\toolname as Zero-Shot Regression Test Generator}
From the perspective of \toolname as a zero-shot regression test generator, we wanted to explore the dependence of WAFLGo's effectiveness on the initial set of seed files (i.e., valid XML, JS, PDF, \review{JSON, and Python} files).
The authors of WAFLGo \cite{Xiang:2024aa} used a sound and fair seed corpus selection strategy for their experiments. \review{We follow its setting to use initial seeds from the UNIFUZZ benchmark \cite{unifuzz}, except for \micropython where we pick seeds from its builtin test suite as UNIFUZZ benchmark does not contain Python seeds.}
For \mujs, \libxml, \poppler, \jerryscript, \review{\jq, and \micropython} there are 19, 14, 100, 19, \review{38 and 13} initial seeds for WAFLGo.
Nevertheless, in \review{ten} cases, the initial corpus already reveals Issues \#166 of \mujs, \#535, \review{\#553, and \#841} of \libxml, \review{and \#3262 of \jq} introduced or fixed in the corresponding commits (cf. Column~\emph{WAFLGo.Init} in Tab.~\ref{tab:clevfuzz}). This raises the question of how close to triggering the corresponding bugs are WAFLGo initial seeds.\vspace{-0.2cm}

\begin{multicols}{2}
  \begin{lstlisting}[caption={Mujs \#65},captionpos=b,language=diff,basicstyle=\ttfamily\scriptsize,label={fig:mujs65}]
 (function() {
 var a = 1, b = 2;
 for(var i = 0; i < 1e5; i++) { /a/g
- if(a === b) {
+ if(a =Error== b) {
   throw new Error;
  }
 })();
\end{lstlisting}

  \columnbreak

  \noindent
  \begin{lstlisting}[caption={Mujs \#145},captionpos=b,language=diff,basicstyle=\ttfamily\scriptsize,label={fig:mujs145}]
 c = 30000;
 a = [];
 for (i = 0; i < 2 * c; i += 1)
   a.push(i%c);
 a.sort(function (x, y) {
   return x - y; });
-print(a[2 * c - 2]);
+print(a[2 * a - 2]);
\end{lstlisting}
\end{multicols}\vspace{-0.4cm}

\textbf{Results}. \autoref{tab:clevfuzz} (Col.~\emph{WAFLGo.Init}) shows the effectiveness of the initial corpus in terms of reaching the code changes (\rmarknospace) and revealing changes in the output (\dmarknospace). Among the \review{50} cases, in addition to the \review{ten (10)} bug-revealing cases, in \review{25} cases, the initial seed corpus at least already reaches the code changes. In two (2) additional cases, only a few characters need to be changed to reveal the bug. Listings~\ref{fig:mujs65} and \ref{fig:mujs145} above show such examples for Issues \#65 and \#145 of \mujs. %
Given this result, we find that \toolname, as a zero-shot regression test generator, also makes an excellent seed generator for regression greybox fuzzers, like WAFLGo.

\result{\textbf{RQ4}.
  While \toolname is substantially faster than WAFLGo, \toolname performs better than WAFLGo in bug reproduction and slightly worse in bug finding.
  However, by upgrading \toolname by fuzzing the generated test cases, our zero-shot approach outperforms WAFLGo, whose performance depends on a user-provided seed corpus, which, as we found, happens to be quite close to bug-revealing already. Hence, we find that \toolname, as a zero-shot regression test generator, also makes an excellent seed generator for
  regression greybox fuzzers.
}

\icse{
  Finally, we elaborate the results versus \emph{undirected}-\toolname, which excludes commit information. We find that the undirected variant significantly under-performs our proposed directed variant (i.e., \review{4} bug versus \review{20} bugs). The effectiveness score drops from \review{1.32} (sometimes output-changing) to \review{0.63} (often not reaching the changes), and the average execution time increases from \review{1} minute to $\sim$\review{3} minutes. These results highlight the importance of commit information for LLM-based regression test generation. Detailed experimental results are provided in Supplementary Material \cite{supplementary}.
}

\section{Threats to Validity}
\label{sec:threats}

\textbf{Construct Validity}.
A key concern is the potential for data leakage and memorization by LLMs, where test set information may inadvertently influence training. In this study, the risk pertains to LLMs memorizing the regression test cases from the WAFLGo benchmark (cf. \autoref{tab:dataset}). \review{To assess this risk, we computed the similarity, in terms of Levenshtein ratio \cite{leven}, between the \toolname-generated test cases and the available bug-triggering test cases that were provided along with the bug report (not the commit).} We find that the majority of \toolname-generated test cases are less than 7\% similar (mean 10\%, max. 40\%) to the ones available online. A significant difference suggests that the LLM did not simply memorize them. We also note that in an experiment where \toolname was asked to generate the command line prompt, as well, a bug was found that---while unrelated to the feature changed in the targeted bug fix---was only fixed after LLM's cut-off date (cf. \textbf{RQ2}).
Our experiment results in \textbf{RQ3} with modified commit message also proves that \toolname does not simply memorize the source code or commit history, as the performance highly depends on the intention. Removing a single word can lead to a significant performance drop.

\textbf{Internal Validity}.
We identified three main threats to internal validity: implementation correctness, confabulation of the LLM, and the randomness of our results. To ensure correctness, we conducted thorough code reviews and testing. Our replication package is made available for transparency. We used publicly accessible APIs of \gpt, \gptmini, and \deepseekr, adhering strictly to their documented usage guidelines. We mitigate the impact of confabulation; all generated test cases are actually executed on subject programs to validate the test outcome. To handle the randomness in our results, we repeated each experiment within the available budget, i.e., ten (10) times, and reported the findings across all trials. %

\textbf{External Validity}.
We do not claim the generality of our results but consider our experiments as an important case study for six open-source C programs that take highly structured, human-readable inputs. We sought to test the capabilities of an LLM as a regression test generation tool and carefully established benchmark selection criteria to align with this goal. We selected \emph{all} programs from the WAFLGo benchmark that apply and \emph{all} of their commits. %
The findings may not extend to programs with less structured or non-human-readable input formats. \review{Finally, our approach does not aim to replace assertion-based regression test suites for well-specified functionality. Instead, it complements them in domains where expected behavior is underspecified or difficult to encode, and where regressions commonly manifest as crashes or semantic inconsistencies.}

\review{\emph{Bug Types and Detection Mechanisms.} Our approach is agnostic to the type of bug being detected and does not rely on memory-specific properties. The only assumption is the availability of an automated mechanism that flags failures when executing generated test cases. In our evaluation, we instantiate this mechanism using ASan to detect memory-related bugs. However, our approach already detects non-memory-related failures in the benchmark. For example, bugs \#5117/5138/5153 in JerryScript are detected via internal assertion failures (ERR\_FAILED\_INTERNAL\_ASSERTION) rather than memory violations. Other detectors, such as developer-provided assertions, runtime checks, differential testing oracles, or semantic consistency checks, could be integrated without changes to the test generation pipeline.}

\section{Related Work}
\label{sec:related_work}
\textbf{LLMs for Automatic Unit Test Generation}.
The field of automated test case generation has evolved significantly, particularly with the advent of deep learning techniques~\cite{atlas,atenatest,codet}.

More recent advancements
have explored the adoption of LLMs for automatic unit test generation.
TESTPILOT~\cite{TESTPILOT} relies on LLMs to generate unit tests by providing the signature, implementation and usage documentation of the function under test, and iteratively refine the tests with execution feedback.
ChatUnitTest~\cite{ChatUniTest} overcomes the prompt context window limit by adaptively selecting the most relevant code elements, and mitigates errors in tests through a generation-validation-repair loop.
SymPrompt~\cite{SymPrompt} generates test inputs for executing a specific path identified during the static analysis.
HITS~\cite{hits} tackles the problem of testing complex methods by using LLMs to decompose method-to-test and generate unit tests slice by slice.
COVERUP~\cite{COVERUP} achieves high coverage by including coverage information in feedback to generate unit tests targeting uncovered code.
These works share the same goal of generating better unit tests to improve overall coverage, while \toolname aims to generate system-level directed input to test specific regression changes. More generally, we are interested in regression test generation as a machine translation task from the intention of a change to a test of that change.

LIBRO~\cite{libro} relies on LLMs to generate and rank candidate test cases from bug reports in structured benchmarks like Defects4J~\cite{Defect4J}.
Testora~\cite{testora} uses LLMs to identify behavioral changes in regression tests generated for API calls from four Python libraries.
Meta also internally deploys LLM-based test generation system~\cite{TestGenLLM,ACH} to find software regression using mutation testing. This system assumes the presence of mutants and leverages them to strengthen regression test suites, while \toolname investigates whether LLMs can independently produce such precise tests in the first place.

\vspace{0.1cm}
\noindent
\toolname is different from previous works in the following ways:%
\begin{itemize}[leftmargin=0.4cm,itemsep=0.05cm,topsep=0.1cm]
  \item It generates system-level input, which can be user-provided and external to the actual code implementation. We demonstrated that LLM can generate effective test cases with lightweight \emph{grey-box} information like an expressive commit message and execution feedback. In contrast, previous works generate unit tests that depends on a specific codebase, requiring heavy \emph{white-box} analysis about the target method (e.g., API signature, full source code);
  \item It is the first evaluation of LLMs with respect to regression test generation. In particular, \toolname focuses on generating bug-revealing test cases from commits while previous works have addressed the \emph{undirected} and the \emph{unit test} generation problem to improve test suite's coverage;
  \item It conducts the first controlled experiments to study the impact of the developer's intention for a change in the form of \emph{natural language} instead of specification on LLM-based test generation, and demonstrates that more specific expressions of intention increases an LLMs effectiveness.
\end{itemize}

Recent works~\cite{intetion,intentionDriven} have also highlighted the role of intention in other software engineering tasks. Guo et al.~\cite{intetion} propose a framework for code refinement, where review comments are translated into structured intentions and then used to guide LLMs in generating code modifications. Similarly, Qi et al.~\cite{intentionDriven} formalizes validation intentions (e.g., objectives, preconditions, and expected results) to generate unit tests. Such approaches focus on extracting and encoding intentions to drive refinement or generation. In contrast, \toolname does not formalize intention, but rather studies how intention expressed in commit messages affects the quality of generated test inputs.

\textbf{LLMs for Fuzzing}.
Several fuzzers leverage LLMs to enhance fuzzing techniques, but each operates with different goals and methods.
ChatAFL~\cite{chatafl} uses LLMs to construct grammars for protocol messages, mutate inputs, and generate sequences to improve fuzzing efficiency, distinguishing it from \toolname, which is not a protocol fuzzer.
PromptFuzz~\cite{PromptFuzz}, meanwhile, utilizes LLMs to iteratively generate fuzz drivers that explore API functions, making it fundamentally different from \toolname, which focuses on fuzzing through input generation rather than fuzz driver creation.
Fuzz4All~\cite{Fuzz4ALL} employs a pure-LLM approach to generate and mutate code snippets testing various compiler features with user-provided documentation and example code.
Jiang et al.~\cite{10.1145/3691620.3695513} conducted the first systematic study comparing LLMs with constraint-based techniques for directed test input generation, evaluating their ability to reach specific target branches.
Cottontail \cite{SP26-cottontail} uses LLMs to extends concolic execution for systematically testing programs taking highly-structured inputs.
TELPA~\cite{REF_TELPA} combines program analysis with LLMs to cover hard-to-cover branches by extracting real usage scenarios and resolving inter-procedural dependencies.
\citet{ICSE26-reachabilityGap} introduce evaluate the utility to test a software system end-to-end to bridge the reachability gap.
In contrast, \toolname generates a few \emph{regression} tests \emph{directed} towards a given commit in a zero-shot manner.

\textbf{Feedback-guided LLM test generation.}
\review{Recent work has confirmed that LLMs require data from external analysis to identify what to test next. MuTAP~\cite{MuTAP} iteratively augments prompts with selected mutants, enabling the model to generate additional tests that exclude remaining mutants and improve mutation score. MUTGEN~\cite{mutgen} extends this idea with a pipeline that relies on mutant analysis, test fixing, and regeneration to maximize fault-detection capability. Panta~\cite{panta} follows the same feedback-directed paradigm, but relies on hybrid program analysis rather than mutation testing: it extracts independent execution paths from a control-flow graph and combines them with dynamic coverage to guide the LLM toward under-tested behaviors. \toolname shares this approach of analysis-guided prompting, but for a different objective: prior works generate unit tests for a single program version with coverage or mutation, while \toolname generates system-level inputs to validate a specific commit, driven by differential execution across versions.}

\section{Discussion}

Throughout the study, we have shown the significant potential of LLMs in generating structured, human-readable, system-level inputs for testing code changes. Our results highlight that even \emph{without} advanced techniques such as fine-tuning or retrieval-augmented generation (RAG), LLMs, when combined with prompting and execution feedback, are capable of producing high-quality test inputs. These inputs can reflect the semantic meaning inside commits. In some cases, they directly trigger the bugs, and in other cases, they could be ``repaired'' to trigger the bugs. So there is potential for LLMs in directed software testing, enabling automated testing workflows that are typically more challenging for conventional tools.

One notable observation is the LLM's ability to generate meaningful input without access to source code context, as long as we describe the intention with accurate natural language. For the philosophical minded, this finding resonates with Wittgenstein's statement in his Tractatus: ``What can be shown cannot be said''. Here what is said is the commit message describing the developer's intention and what is shown is the concrete executable test demonstrating the consequence of that intention.
Our results show that the inverse might be true for automated software engineering: much of what can be ``said'' can indeed be ``shown''. Our work unlocks the potential of ``vibe testing'' as a new form of ``language game''~\cite{LanguageGame}, where users describe high-level intention with natural language, and LLM-powered systems instantiate the meaning into concrete tests to trigger specific behavior of the underlying program, with this ``vibe'' becoming a tangible, testable reality.

While using our tool \toolname standalone has already shown promising results in generating effective test cases in a short time, we believe the value of \toolname shines even more in their subsequent use as a tool in a developer's hand. The LLM-generated test cases are easily comprehensible by human developers, as they are structured and human-readable; we demonstrated how human developers can modify these test cases to trigger bugs. They are also an excellent seed generator for regression greybox fuzzers; even a vanilla greybox fuzzer can perform similarly or better than WAFLGo, which requires a user-provided seed corpus.

\section{Data Availability}

The implementation code, data, and scripts used in this study are available at \url{https://github.com/niMgnoeSeeL/cleverest}.

\begin{acks}
We thank the anonymous reviewers for their constructive feedback and for helping us improve this paper.
We thank Yi Xiang, the author of WAFLGo paper, for answering questions about running WAFLGo.
This research is partially funded by the European Union. Views and opinions expressed are however those of the author(s) only and do not necessarily reflect those of the European Union or the European Research Council Executive Agency. Neither the European Union nor the granting authority can be held responsible for them. Besides funding this work, the funding source had no involvement in the conduct of the research and in the preparation of the article.
This work is supported by ERC grant (Project AT\_SCALE, 101179366).
It is also partially funded by NextGenerationEU and by the 2023 STARS Grants@Unipd programme (Acronym and title of the project: ``PatchThemAll: A Virtualization-Based Approach for Distributing Android Security Patches on Any Custom Android OS'').
It is also partially funded by the Deutsche Forschungsgemeinschaft (DFG, German Research Foundation) under Germany's Excellence Strategy - EXC 2092 CASA - 390781972.
\end{acks}

\bibliographystyle{ACM-Reference-Format}
\bibliography{ref}

\end{document}